\documentclass[aps,preprint,onecolumn,secnumarabic,nobalancelastpage,amsmath,amssymb,
nofootinbib]{revtex4}

\usepackage{graphics}      % standard graphics specifications
\usepackage{graphicx}      % alternative graphics specifications
\usepackage{longtable}     % helps with long table options
\usepackage{pst-plot}
\usepackage{dcolumn}% Align table columns on decimal point
\usepackage{hyperref}% add hypertext capabilities
\usepackage{bm}            % special 'bold-math' package
\usepackage{mathrsfs}

\begin{document}

\preprint{APS/123-QED}

\title{Study of the geodesic equations of a spherical symmetric spacetime in conformal gravity}

\author{Bahareh Hoseini}
\author{Reza Saffari}%
\email{rsk@guilan.ac.ir}
\author{Saheb Soroushfar}

\affiliation{Department of Physics, University of Guilan, 41335-1914, Rasht, Iran.
%${}^2$Institut f\"ur Physik, Universit\"at Oldenburg, Postfach 2503 D-26111 Oldenburg, Germany.
}

\date{\today}

\begin{abstract}
 Set of analytic solutions of the geodesic equation in a spherical conformal
spacetime is presented. Solutions of this geodesics
can be expressed in terms of the Weierstrass $\wp$ function and the Kleinian $\sigma$ function. Using
conserved energy and angular momentum we can characterize the different orbits.
Also, considering parametric diagrams and effective potentials, we plot some possible orbits. Moreover, with the help of analytical solutions, we investigate the light deflection for such an escape orbit.
\end{abstract}

\maketitle

\section{INTRODUCTION}
%%%%%%%%%%%%%%%%%%%%%%%%%%%%%%
One of the alternative theory to standard model of General Relativity is Conformal Gravity (CG),
which is able to solve some problems of Einstein gravity. Thus it is worth investigating this theory in more detail. CG (see \cite{Mannheim:2005bfa} and references therein) can produce a linear terms  which explain galactic rotation curve without considering dark matter, and it confirms Newtonian gravity in solar system. Also it explains the accelerated expansion of the universe without dark energy. 
This theory is in four dimensions  and the action of CG consists of the Weyl tensor, $S_{conf} =\int d^4x \sqrt{g}W^2$. The conformal transformation is $g_{\mu\nu}\rightarrow \Omega^2(x)g_{\mu\nu}$, which sensitive to the angle, but not to distances. Study of the orbits of test particles and light rays in conformal spacetimes help to understand
the physical properties of Conformal field equations.  
 On the other hand, the motion of matter and light can be used to classify a given spacetime, and to highlight its characteristics. Using the analytic solutions of the geodesic equation, we obtain some information about the spacetime properties of the black holes  \cite{Hioki:2009na}.
% gravitational waves models for Extreme Mass Ratio Inspirals (EMRI) \cite{Barack:2006pq}, motion of effective one–body \cite{Damour:1999cr}, finding homoclinic orbits \cite{PerezGiz:2008yq}, testing numerical codes for binary systems and finally may also be of use for practical applications like geodesy.
In 1931 Hagihara \cite{Y. Hagihara}, solved the geodesic equation  
in a Schwarzschild spacetime using applied of the  elliptic Weierstrass function.
The solutions for Kerr and Kerr-Newman space times have the same mathematical structure \cite{S. Chandrasekhar} and can be solved analogously. The mathematical method which is used to solve the equations of motion in Schwarzschild (anti) de Sitter spacetime, is the hyper elliptic functions. These functions are based on the solution of the Jacobi inversion problem. \cite{Hackmann:2008zza,Hackmann:2008zz}. Also these methods are applicabied for other spacetime such as, Reissner-Nordstrom and Reissner-Nordstrom -(anti)-de Sitter \cite{Hackmann:2008tu} and spinning black ring spacetimes \cite{Grunau:2012ai}.
 Moreover, this analytically method are used for some spacetimes such as $f(R)$ gravity, BTZ, GMGHS black holes, cylindrically symmetric conformal and (rotating) black string-(anti-) de sitter in Refs.~ \cite{Soroushfar:2015wqa,Soroushfar:2015dfz,Soroushfar:2016yea,Hoseini:2016nzw,Soroushfar:2016esy,Soroushfar:2016azn,Kazempour:2016dco}. Here, we study geodesic equations for spherically symmetric in the vicinity of blackhole in conformal gravity. In CG, some investigations predicate that the only null geodesics are 
physically meaningful in the theory of conformal gravity \cite{Brihaye:2009ef,Barabash:1999bj,Wood:2001ve}.
 But, there are some different descriptions on applicability of geodesic solution in this theory,
% but there are some discussions on applicability of timelike geodesics in conformal gravity, 
such as, free  fall of elementary particles or packets of gravitational energy of geons 
which is proposed by Wheeler \cite{Ohanian:2015wva}. Also, CG theory for all geodesics makes attractive by choosing the different gauge \cite{Edery:2001at}. 
% Also the possibility of choosing  the gauge which 
%makes attractive the CG theory for all geodesics is described by Edery and $et.al$ \cite{Edery:2001at}. 
One of other applications of geodesic solutions in conformal gravity, is 
the effect of the linear term in the metric on perihelion shift in the time like geodesics, 
which is studied by Sultana and $et.al$ \cite{Sultana:2012qp}. We will discuss the geodesic motion of  test particles and light rays in conformal gravity with a spherical symmetry. Our paper is organized as follows:
In Sec.~\ref{FE}, we start by studying properties of the spherical spacetime in conformal gravity. In Sec.~\ref{GE} we obtain the geodesic equations. In Sec.~\ref{ASE} 
the geodesic equation solve in terms of Weierstrass elliptic function in the case of null geodesic , and derivatives of Kleinian sigma functions in the case of timelike geodesic. Also, we discuss about the light deflection of fly by orbit type for null geodesic. We plot some possible orbits 
in Sec.~\ref{O} and we conclude in Sec.~\ref{C}.
%%%%%%%%%%%%%%%%%%%%%%%%%%%%%%%%

\section{SPHERICAL SOLUTION IN CONFORMAL WEYL GRAVITY}\label{FE}
In this section, we study the field equation and metric in conformal gravity.
The conformal invariance leads to the unique action 
\begin{equation}\label{action}
I_W=-\alpha \int d^4x(-g)^{1/2}C_{\kappa\lambda\mu\nu}C^{\kappa\lambda\mu\nu} 
\end{equation}
\begin{equation}\label{C}
C_{\kappa\lambda\mu\nu}=R_{\kappa\lambda\mu\nu}-\frac{1}{2}(g_{\kappa\mu}R_{\lambda\nu}-g_{\kappa\nu}R_{\lambda\mu}+g_{\lambda\nu}R_{\kappa\mu}-g_{\lambda\mu}R_{\kappa\nu})+\frac{R}{6}(g_{\kappa\mu}g_{\lambda\nu}-g_{\kappa\nu}g_{\lambda\mu}),
\end{equation}
where $C_{\kappa\lambda\mu\nu}$ is the conformal Weyl tensor and $\alpha$ is a
purely dimensionless coefficient. The action of Eq.~(\ref{action}) leads to the following gravitational field equations
\begin{equation}\label{bach}
2 \alpha_g W_{\mu\nu}=\frac{1}{2}T_{\mu\nu}.
\end{equation}
in the presence of a energy momentum tensor and with $W_{\mu\nu}$ being given by
\begin{equation}
W_{\mu\nu}=\frac{1}{3}\nabla_\mu \nabla_\nu R-\nabla_\lambda \nabla^\lambda R_{\mu\nu}+\frac{1}{6}(R^2+\nabla_\lambda \nabla^\lambda R-3R_{\kappa\lambda}R^{\kappa\lambda})g_{\mu\nu} +2R^{\kappa\lambda}R_{\mu\kappa\nu\lambda}-\frac{2}{3}RR_{\mu\nu}
\end{equation} 
The exact static and spherically symmetric vacuum solution for conformal gravity is given, up to a conformal factor, by the metric 
\begin{equation}\label{metric}
ds^2=-B(r)dt^2+\frac{dr^2}{B(r)} +r^2(d\theta^2 +sin^2\theta d\varphi^2),
\end{equation}
where 
\begin{equation}\label{metrics}
B(r)=1-\frac{\beta(2-3\beta\gamma)}{r}-3\beta\gamma +\gamma r-kr^2,
\end{equation}
and $\beta$, $\gamma$, and $k$ are integration constants. The parameter $\gamma$ measures the departure of Weyl theory from general relativity, and for small enough $\gamma$, both theories have similar predictions. In Eq.~(\ref{metrics}) $k$ acts like a cosmology constant, and $\beta$ is related to mass \cite{Mannheim:1988dj,Mannheim:1990ya}.

%%%%%%%%%%%%%%%%%%%%%%%%%%%
% Next section
%%%%%%%%%%%%%%%%%%%%%%%%%%%
\section{THE GEODESIC EQUATION}\label{GE}
In this section we derive geodesic equations for particles and light rays.  
The general form of geodesic equation is
\begin{equation}\label{geodesy}
\frac{d^2x^\mu}{ds^2}+\Gamma^\mu_{\rho\sigma}\frac{dx^\rho}{ds}\frac{dx^\sigma}{ds} =0,
\end{equation}
where
$\Gamma^\mu_{\rho\sigma}$ are the Christoffel symbols.
The conserved energy and the angular momentum are
\begin{equation}\label{E}
E=-g_{tt}\frac{dt}{ds}=\frac{dt}{ds}(1 - \frac{\beta(2 - 3 \beta \gamma)}{r} - 3\beta\gamma+\gamma r-kr^2),
\end{equation}
\begin{equation}\label{L}
L=g_{\varphi\varphi}\frac{d\varphi}{ds}=r^2\frac{d\varphi}{ds} \, .
\end{equation}
Using the normalization condition
$ds^2=\frac{1}{2}g_{\mu\nu}\frac{dx^\mu}{ds}\frac{dx^\nu}{ds}=-\frac{1}{2}\epsilon$, and from Eq.~(\ref{geodesy}) we obtain
equations for $r$ as a functions of $\tau,\phi,t$ 
\begin{equation}\label{rtu}
(\frac{dr}{d\tau})^2=E^2-(1 - (\frac{\beta(2 - 3 \beta \gamma)}{r} - 3\beta\gamma+\gamma r-kr^2)(\epsilon +\frac{L^2}{r^2}),
\end{equation}
\begin{equation}\label{rphi}
(\frac{dr}{d\phi})^2=\frac{r^4}{L^2}(E^2-(1-\frac{\beta(2-3\beta\gamma)}{r}-3\beta\gamma+\gamma r -kr^2)(\epsilon +\frac{L^2}{r^2}))=R(r),
\end{equation}
\begin{equation}\label{rt}
(\frac{dr}{dt})^2=\frac{1}{E^2}(1-\frac{\beta(2-3\beta\gamma)}{r}-3\beta\gamma+\gamma r -kr^2)^2(E^2-(1-\frac{\beta(2-3\beta\gamma)}{r}-3\beta\gamma+\gamma r -kr^2)(\epsilon +\frac{L^2}{r^2})),
\end{equation}
where for massive particles $\epsilon=1$ and for light $\epsilon=0$.
Eq.~(\ref{rtu}) suggests the introduction of an effective potential
\begin{equation}\label{veff}
V_{eff}=(1-\frac{\beta(2-3\beta\gamma)}{r}-3\beta\gamma+\gamma r -kr^2)(\epsilon +\frac{L^2}{r^2}).
\end{equation}
The shapes of effective potential are in Figs.~(\ref{V1o0}--\ref{Vnull2z}).

\section{Analytical solution of geodesic equations}\label{ASE}

In this section, using elliptic Weierstrass $\wp$-function, and Kleinian $\sigma$ function (for both test particles and light rays) we obtain the analytical solutions of geodesic equations in conformal spacetime Eq.~(\ref{metric}). Also we use this solutions to plot the possible orbits for light and test particle around this spacetime.

\subsection{The $\tilde{r}$-$\phi$-equation}\label{req}
For the analysis of the parameters of the spacetime we use dimensionless quantities, as follow
\begin{equation}
\tilde{r}=r/M, \tilde{\beta}=\beta/M, \tilde{\gamma}=M\gamma, \tilde{k}=kM^2, \mathcal{L}=M^2/L^2
\end{equation}
Thus, Eq.~(\ref{rphi}) can be written as
\begin{eqnarray}\label{Rtild}
(\frac{d\tilde{r}}{d\phi})^2 &=&\tilde{k}\epsilon \mathcal{L}\tilde{r}^6-\tilde{\gamma}\epsilon\mathcal{L}\tilde{r}^5+(k+3\epsilon\mathcal{L}\tilde{\beta}\tilde{\gamma}-\epsilon \mathcal{L}+E^2\mathcal{L})\tilde{r}^4 \nonumber \\
&-&(\epsilon \mathcal{L} \tilde{\beta} (2-3\tilde{\beta}\tilde{\gamma})-\tilde{\gamma})\tilde{r}^3  -(3\tilde{\beta}\tilde{\gamma}-1)\tilde{r}^2\nonumber  \\
&-&\tilde{\beta}(2-3\tilde{\beta}\tilde{\gamma})\tilde{r}=R(\tilde{r}).
\end{eqnarray}
In following, we solve equation (\ref{Rtild}) analytically.
\subsubsection{Null geodesics}
In this part, we consider null geodesics and solve Eq.~(\ref{Rtild}). For $\epsilon=0$, Eq.~(\ref{Rtild}) becomes to polynomial of degree four and it can be reduced to third order by substituting $\tilde{r}=\frac{1}{u}$
\begin{eqnarray}\label{p3}
(\frac{du}{d\varphi})^2&=&-\tilde{\beta}(2-3\tilde{\beta}\tilde{\gamma})u^3-(3\tilde{\beta}\tilde{\gamma}-1)u^2-\epsilon \mathcal{L} \tilde{\beta} ((2-3\tilde{\beta}\tilde{\gamma})-\tilde{\gamma})u\nonumber  \\
&+&(k+3\epsilon\mathcal{L}\tilde{\beta}\tilde{\gamma}-\epsilon \mathcal{L}+E^2\mathcal{L})=P_3(u)=\sum_{i=0}^3 a_i u^i.
\end{eqnarray}
With next substitution
\begin{equation}
u=\frac{1}{a_3}(4y-\frac{a_{2}}{3})=\frac{-1}{\tilde{\beta}(2-3\tilde{\beta}\tilde{\gamma})}(4y+\frac{1}{3}(3\tilde{\beta}\tilde{\gamma}-1)),
\end{equation}
 $P_3(u)$ transforms in to the Weierstrass form, so that Eq.~(\ref{p3}) turns into:
\begin{equation}\label{p31y}
(\frac{dy}{d\varphi})^2=4y^3-g_2y-g_3=P_3(y),
\end{equation}
with
\begin{equation}
g_2=\frac{a_2^2}{12}-\frac{a_1a_3}{4}=\frac{1}{12}(3\tilde{\beta}\tilde{\gamma}-1)^2-\frac{1}{4}\epsilon \tilde{\beta} \mathcal{L}((2-3\tilde{\beta}\tilde{\gamma})-\tilde{\gamma})(\tilde{\beta}(2-3\tilde{\beta}\tilde{\gamma})),
\end{equation}
\begin{eqnarray}
g_3&=&\frac{a_1a_2a_3}{48}-\frac{a_0a_3^2}{16}-\frac{a_2^3}{216}=\frac{1}{216}(3\tilde{\beta}\tilde{\gamma}-1)^3\nonumber  \\
&-&\frac{1}{48}\epsilon \tilde{\beta} \mathcal{L}((2-3\tilde{\beta}\tilde{\gamma})-\tilde{\gamma})(3\tilde{\beta}\tilde{\gamma}-1)\tilde{\beta}(2-3\tilde{\beta}\tilde{\gamma})\nonumber  \\
&-&\frac{1}{16}(k+3\epsilon\mathcal{L}\tilde{\beta}\tilde{\gamma}-\epsilon \mathcal{L}+E^2\mathcal{L})(\tilde{\beta}(2-3\tilde{\beta}\tilde{\gamma}))^2.
\end{eqnarray}
Eq.~(\ref{p31y}) is of elliptic type and solved by the Weierstrass function \cite{Hackmann:2008zz,M.Abramowitz,E. T. Whittaker},
\begin{equation}\label{wp}
y(\varphi)=\wp(\varphi-\varphi_{in};g_2,g_3),
\end{equation}
where, $\varphi_{in}=\varphi_0+\int_{y_0}^\infty \frac{dy}{\sqrt{4y^3-g_2y-g_3}}$,
with
\begin{equation}
y_0=\frac{a_3}{4\tilde{r}_0}+\frac{a_2}{12}=-\frac{1}{4\tilde{r}_0}(\tilde{\beta}(2-3\tilde{\beta}\tilde{\gamma}))-\frac{1}{12}(3\tilde{\beta}\tilde{\gamma}-1).
\end{equation}
Then, the solution of Eq.~(\ref{Rtild}) in the case of $\epsilon=0$, acquires the form
\begin{equation}
\tilde{r}(\varphi)=\frac{a_3}{4\wp(\varphi-\varphi_{in};g_2,g_3)-\frac{a_2}{3}}=\frac{-(\tilde{\beta}(2-3\tilde{\beta}\tilde{\gamma}))}{4\wp(\varphi-\varphi_{in};g_2,g_3)+\frac{1}{3}(3\tilde{\beta}\tilde{\gamma}-1)}.
\end{equation}

One application of this analytical solution is calculated the deflection angle. 
The following expression was presented for the deflection angle, by Rindler and Ishak in Ref(\cite{Rindler:2007zz})
\begin{equation}
\tan\psi=\frac{B(r)^{\frac{1}{2}}r}{dr/d\varphi}.
\end{equation}  
By replacing the expression $k+\frac{E^2}{L^2}$ in Eq.~(\ref{rphi}) with the equation in terms of $r_p$, the angle of deviation is given by
\begin{equation}\label{light}
\tan\psi= \frac{\sqrt{1-\frac{\beta (2-3\beta\gamma)}{r}-3\beta\gamma +\gamma r-kr^2}}{\sqrt{\frac{r^2}{r_p^2}(3\beta\gamma -1+\frac{\beta(2-3\beta\gamma)}{r_p}-\gamma r_p)-(3\beta\gamma -1+\frac{\beta(2-3\beta\gamma)}{r}-\gamma r)}}.
\end{equation}
The Eq.~(\ref{light}) is valid for all light rays, not only for a small
deflection as discussed in \cite{Cattani:2013dla}.

\subsubsection{Timelike geodesics}\label{ssTg}
Considering the case $\epsilon=1$, Eq.~(\ref{Rtild}),  is of hyperelliptic type. Using the substitution $\tilde{r}=\frac{1}{u}$ it can be rewritten as
\begin{eqnarray}\label{p5}
(u\frac{du}{d\varphi})^2&=&-\tilde{\beta}(2-3\tilde{\beta}\tilde{\gamma})u^5-(3\tilde{\beta}\tilde{\gamma}-1)u^4-(\epsilon \mathcal{L} \tilde{\beta}(2-3\tilde{\beta}\tilde{\gamma})-\tilde{\gamma})u^3\nonumber  \\
&+&(k+3\epsilon\mathcal{L}\tilde{\beta}\tilde{\gamma}-\epsilon \mathcal{L}+E^2\mathcal{L})u^2-\frac{\tilde{\gamma}\epsilon\mathcal{L}}{4}u+\tilde{k}\epsilon \mathcal{L}\nonumber  \\
&=&P_5(u)=\sum_{i=0}^5 a_i u^i.
\end{eqnarray}

This problem is a special case of the Jacobi inversion problem and can be solved when restricted to the $\theta$ divisor, the set of zeros of the Riemann $\theta$ function. The solution procedure is extensively discussed in Refs.~\cite{Hackmann:2008zz,Enolski:2010if}. The analytic solution of Eq.~(\ref{p5}) is given in terms of derivatives of the Kleinian $\sigma$ function
\begin{equation}
	u(\varphi) = \left. \frac{\sigma_1 (\boldsymbol{\varphi}_\infty)}{\sigma_2 (\boldsymbol{\varphi}_\infty)} \right| _{ \sigma (\boldsymbol{\varphi}_\infty)=0} \, ,
\end{equation}
with
\begin{equation}
	\boldsymbol{\varphi}_\infty = 
	\left( 
	\begin{array}{c}
	  \varphi_2 \\
   	  \varphi-\varphi_{\rm in}'
	\end{array}
	\right),
\end{equation}
and $ \varphi_{\rm in}'=  \varphi_{\rm in}+\int_{ \varphi_{\rm in}}^{\infty}\! \frac{u \, \mathrm{d} u'}{\sqrt{ P_5(u')}}$. The component $ \varphi_2$ is specified by the condition $\sigma (\boldsymbol{\varphi}_\infty)=0$.
The function $\sigma_i$ is the $i$-th derivative of Kleinian $\sigma$
function and $\sigma_z$ is
\begin{equation}
\sigma(z)=Ce^{zt}kz \theta [g,h](2\omega^{-1}z;\tau),
\end{equation}
which is given by the Riemann $\theta$-function with characteristic $[g,h]$. $(2\omega, 2\acute{\omega})$ is the period-matrix, $\tau$ is the symmetric Riemann matrix, $(2\eta, 2\acute{\eta})$ is the periodmatrix of the second kind , $\kappa = \eta(2ω)^{-1}$ is the matrix 
and $2[g,h] = (0, 1)^t+(1,1)^t\tau$ is the vector of Riemann
constants with base point at infinity. The constant $C $, can be
given explicitly, see e.g.\cite{V. M. Buchstaber}, but does not matter here. 

Finally the analytical solution of Eq.~(\ref{Rtild}) is
\begin{equation}
r(\varphi) = \left. \frac{\sigma_2 (\boldsymbol{\varphi}_\infty)}{\sigma_1 (\boldsymbol{\varphi}_\infty)}\right| _{ \sigma (\boldsymbol{\varphi}_\infty)=0} \, .
\end{equation}
This is the analytic solution of the equation of motion
of a test particle in spherical
space time in coformal gravity. The solution is valid in all regions of this spacetime.

\section{ORBITS}\label{O}
\begin{figure}[ht]
\centerline{\includegraphics[width=8cm]{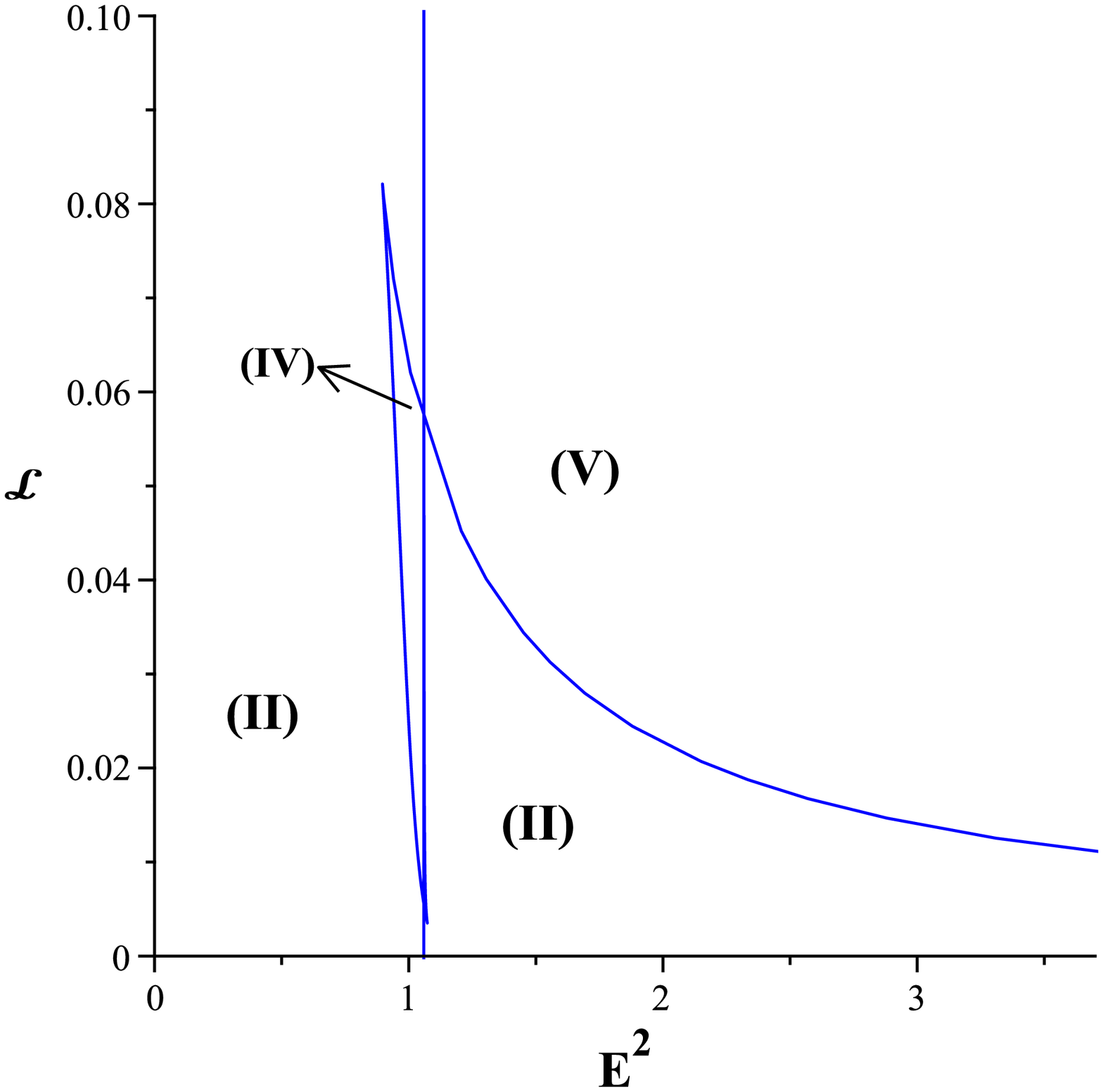}}
\caption{\label{el1}\small   
 Parametric $\mathcal{L}$-$E^2$-diagram with the parameters $\varepsilon=1,\tilde{\beta}=1,\tilde{\gamma}=10^{-3}, \tilde{k}=(\frac{1}{3})10^{-5}$. There are tree different regions. $R^{*}$ has no zero in region V, two zeros in region II and four zeros in region IV  for conformal gravity.
}
\end{figure}

\begin{figure}[ht]
\centerline{\includegraphics[width=8cm]{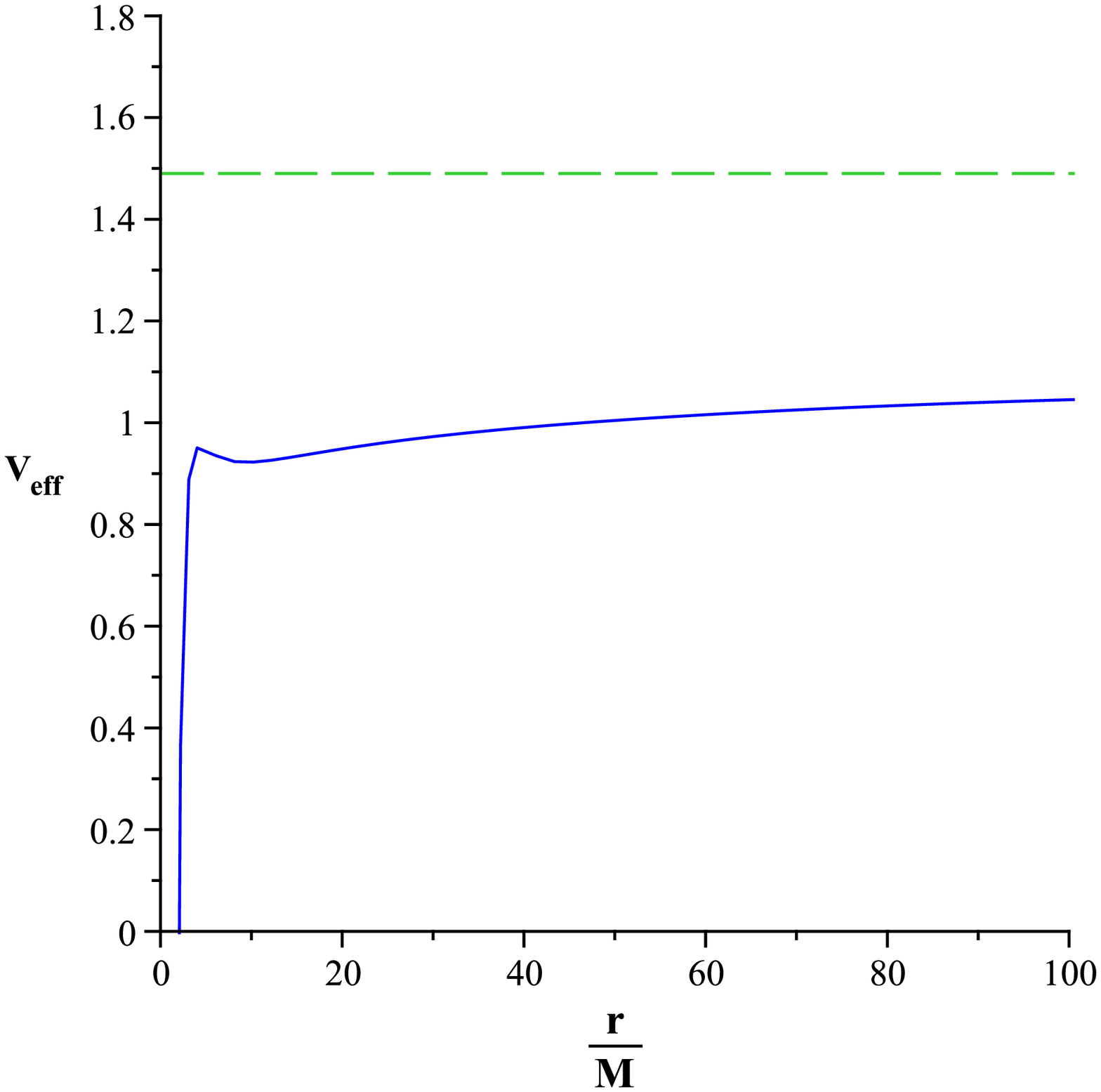}}
\caption{\label{V1o0}\small   
 Effective potentials for region V of parametric diagram in Fig.~(\ref{el1}) with the parameters $\mathcal{L}=0.07, \varepsilon=1,\tilde{\beta}=1,\tilde{\gamma}=10^{-3},\tilde{k}=(\frac{1}{3})10^{-5}$. The horizontal green dashed
line represents the squared energy parameter $E^2=1.5$.
}
\end{figure}

\begin{figure}[ht]
\centerline{\includegraphics[width=8cm]{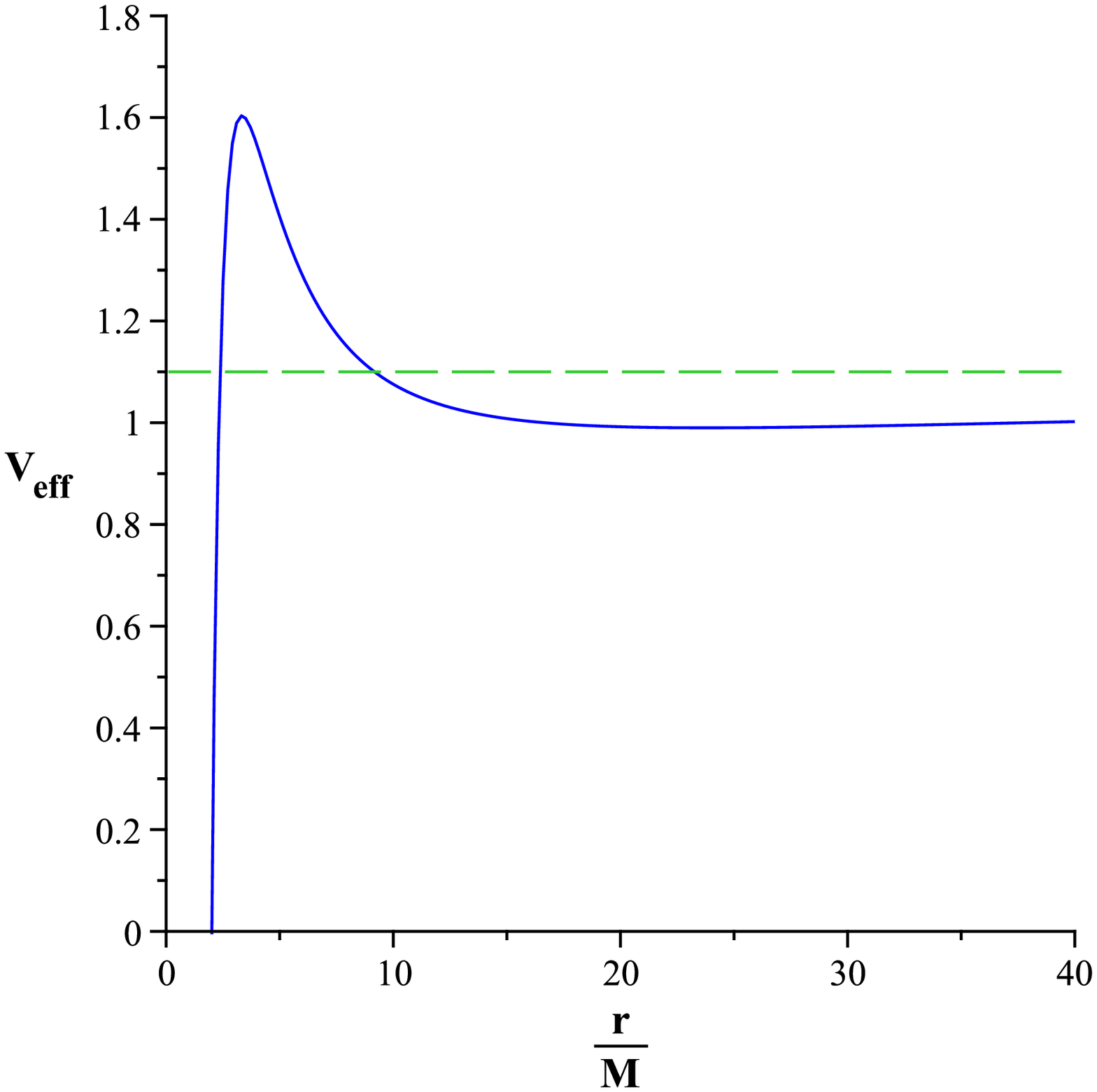}}
\caption{\label{V1o2}\small   
Effective potentials for region II of parametric diagram in Fig.~(\ref{el1}) with the parameters $\mathcal{L}=0.03, \varepsilon=1,\tilde{\beta}=1,\tilde{\gamma}=10^{-3},\tilde{k}=(\frac{1}{3})10^{-5}$. The horizontal green dashed
line represents the squared energy parameter $E^2=1.1$. 
}
\end{figure}

\begin{figure}[ht]
\centerline{\includegraphics[width=8cm]{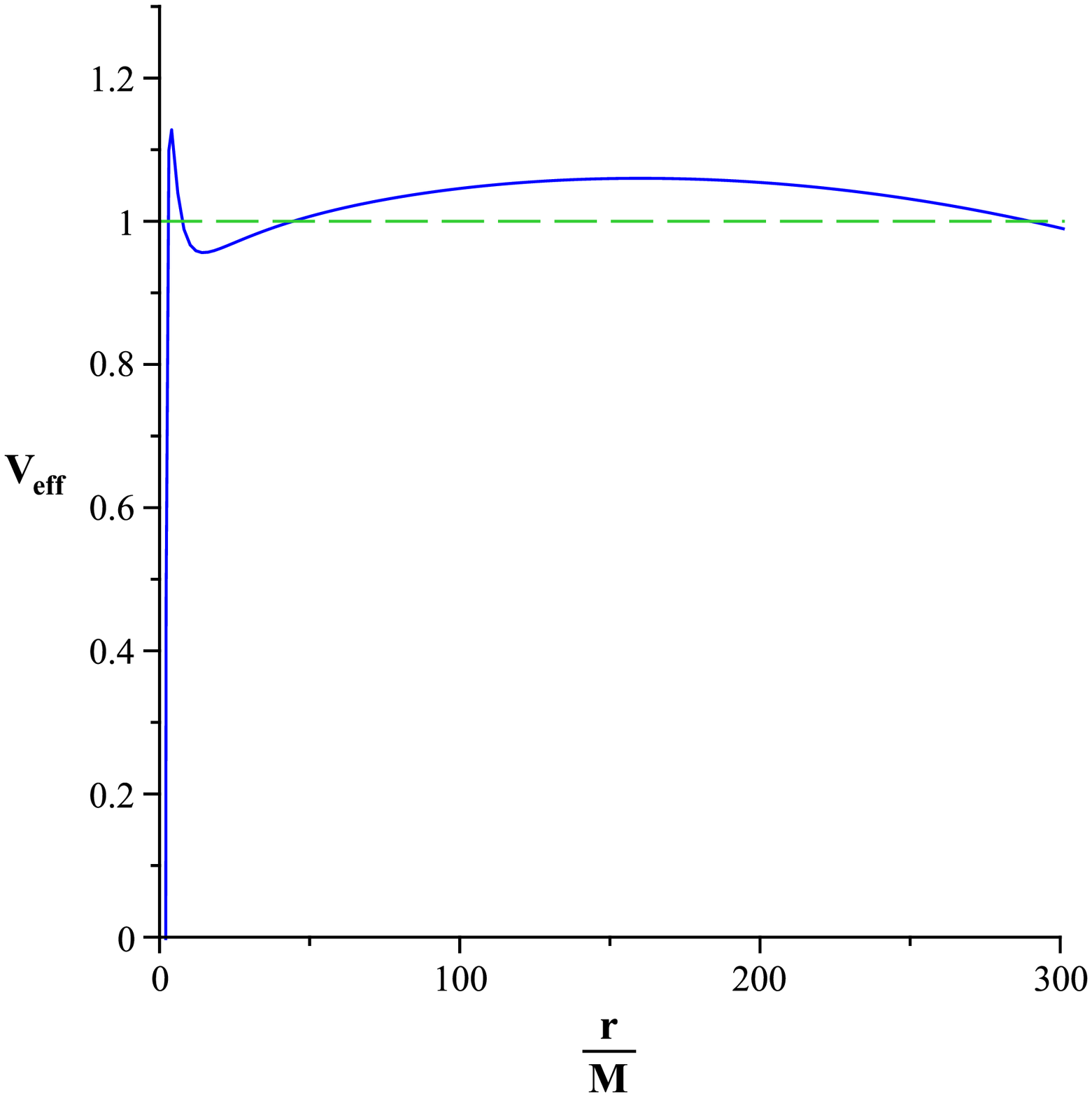}}
\caption{\label{V1o4}\small   
Effective potentials for region IV of parametric diagram in Fig.~(\ref{el1}) with the parameters $\mathcal{L}=0.05, \varepsilon=1,\tilde{\beta}=1,\tilde{\gamma}=10^{-3},\tilde{k}=(\frac{1}{3})10^{-5}$. The horizontal green dashed
line represents the squared energy parameter $E^2=1$.
}
\end{figure}

In this section, we analyze the possible orbits by help of these analytical solution, parametric diagram Fig.~(\ref{el1}) and effective potential Figs.~(\ref{V1o0}--\ref{V1o4}). 
Eq.~(\ref{Rtild}) implies that $R(\tilde{r})\geq 0$ is a necessary condition for the existence of a geodesic motion. The real and positive zeros of $R(\tilde{r})$, that determine the possible types of orbits, are the turning points of the geodesics. 
some possible orbits, which may occur in this spacetime are
\begin{enumerate}
\item  \emph{Escape orbits} (EO): the particle approaches the black hole and then escape its gravity.
\item  \emph{Bound orbits} (BO): the particle moves between two points.
\item  \emph{Terminating bound orbits} (TBO): particle motion end in the singularity at $\tilde{r}=0$
\item  \emph{Terminating escape orbits} (TEO): particle motion start from infinity and end in the singularity at $\tilde{r}=0$. 
\end{enumerate}

\begin{figure}[ht]
\centerline{\includegraphics[width=8cm]{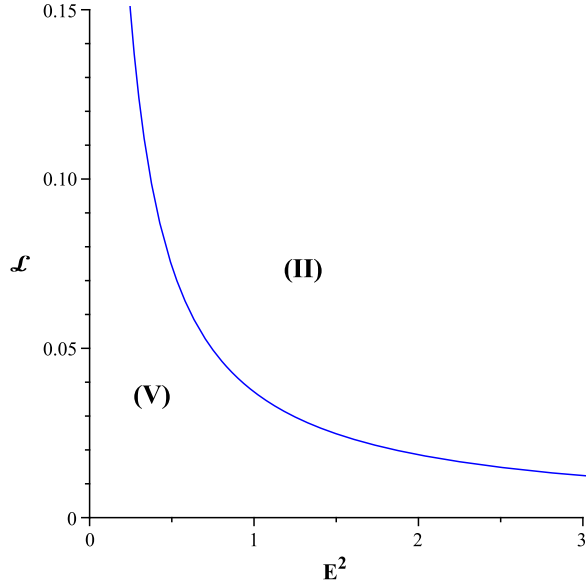}}
\caption{\label{elnull}\small   
 Parametric $\mathcal{L}$-$E^2$ diagram with the parameters $\varepsilon=0,\tilde{\beta}=1,\tilde{\gamma}=10^{-3}, \tilde{k}=\frac{10^{-5}}{3}$. There are two different regions. $R^{*}$ has none zero in region I, two zeros in region II, for conformal gravity.
}
\end{figure}

\begin{figure}[ht]
\centerline{\includegraphics[width=8cm]{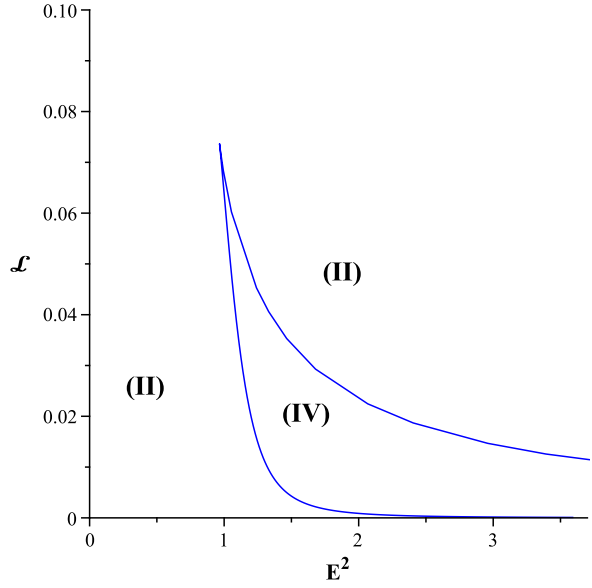}}
\caption{\label{elgama2}\small   
 Parametric $\mathcal{L}$-$E^2$-diagram with the parameters $\varepsilon=0,\tilde{\beta}=1,\tilde{\gamma}=10^{-2}, \tilde{k}=(\frac{1}{3})10^{-5}$. There are three different regions. $R^{*}$ has four zeros in region IV and two zeros in region II, for conformal gravity.
}
\end{figure}

\begin{figure}[ht]\label{ELgama0}
\centerline{\includegraphics[width=8cm]{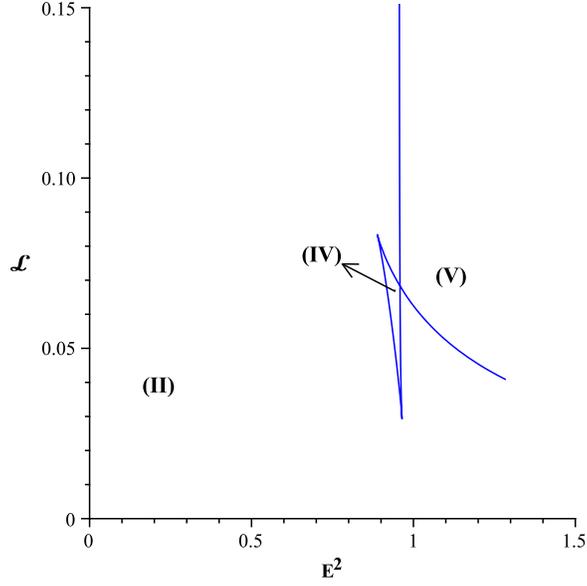}}
\caption{\label{elshdesiter}\small   
 Parametric $\mathcal{L}$-$E^2$-diagram with the parameters $\varepsilon=1,\tilde{\beta}=1,\tilde{\gamma}=0, \tilde{k}=(\frac{1}{3})10^{-5}$. There are three different regions. $R^{*}$ has no zero in region V, two zeros in region II, four zeros in region IV, for Schwarzschild-de Sitter spacetime.
}
\end{figure}

\begin{figure}[ht]\label{ELlanda0}
\centerline{\includegraphics[width=8cm]{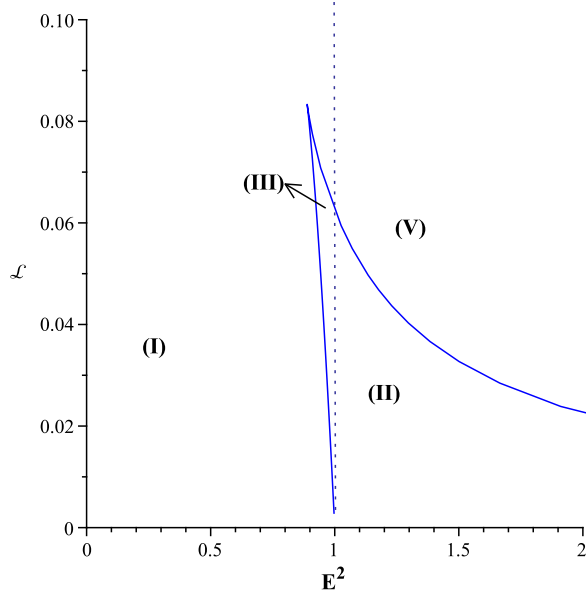}}
\caption{\label{elsh}\small   
 Parametric $\mathcal{L}$-$E^2$-diagram with the parameters $\varepsilon=1,\tilde{\beta}=1,\tilde{\gamma}=0, \tilde{k}=0$. There are tree different regions. $R^{*}$ has one zero in region I, two zeros in region II, none zero in region III and four zeores in region IV for Schwarzschild spacetime.
}
\end{figure}

\begin{figure}[ht]
\centerline{\includegraphics[width=8cm]{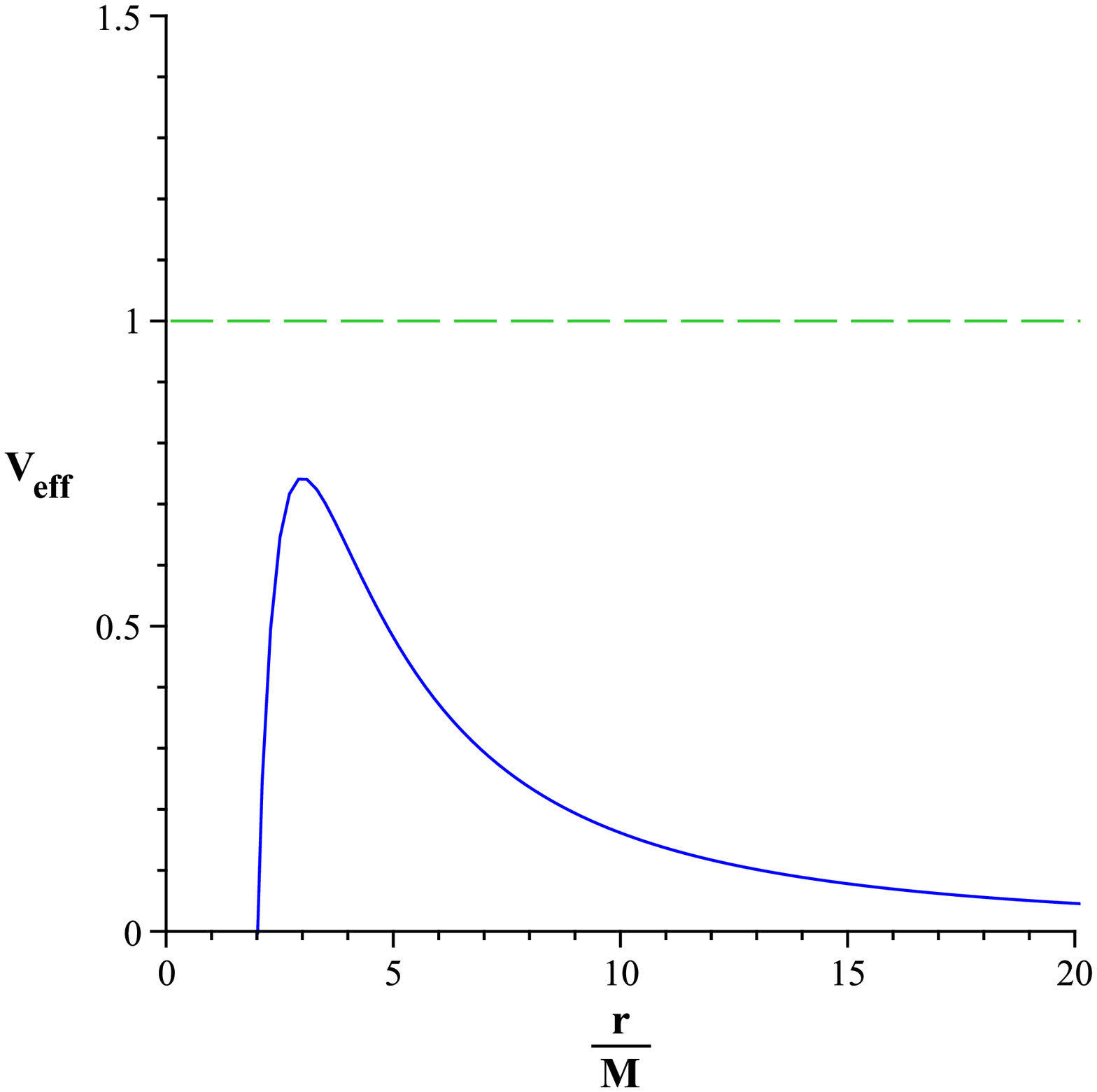}}
\caption{\label{Vnull0z}\small   
 Effective potentials for region V of parametric diagram in Fig.~(\ref{elnull}) with the parameters $\mathcal{L}=0.05, \varepsilon=0,\tilde{\beta}=1,\tilde{\gamma}=10^{-3},\tilde{k}=(\frac{1}{3})10^{-5}$. The horizontal green dashed
line represents the squared energy parameter $E^2=1$.
}
\end{figure}

\begin{figure}[ht]
\centerline{\includegraphics[width=8cm]{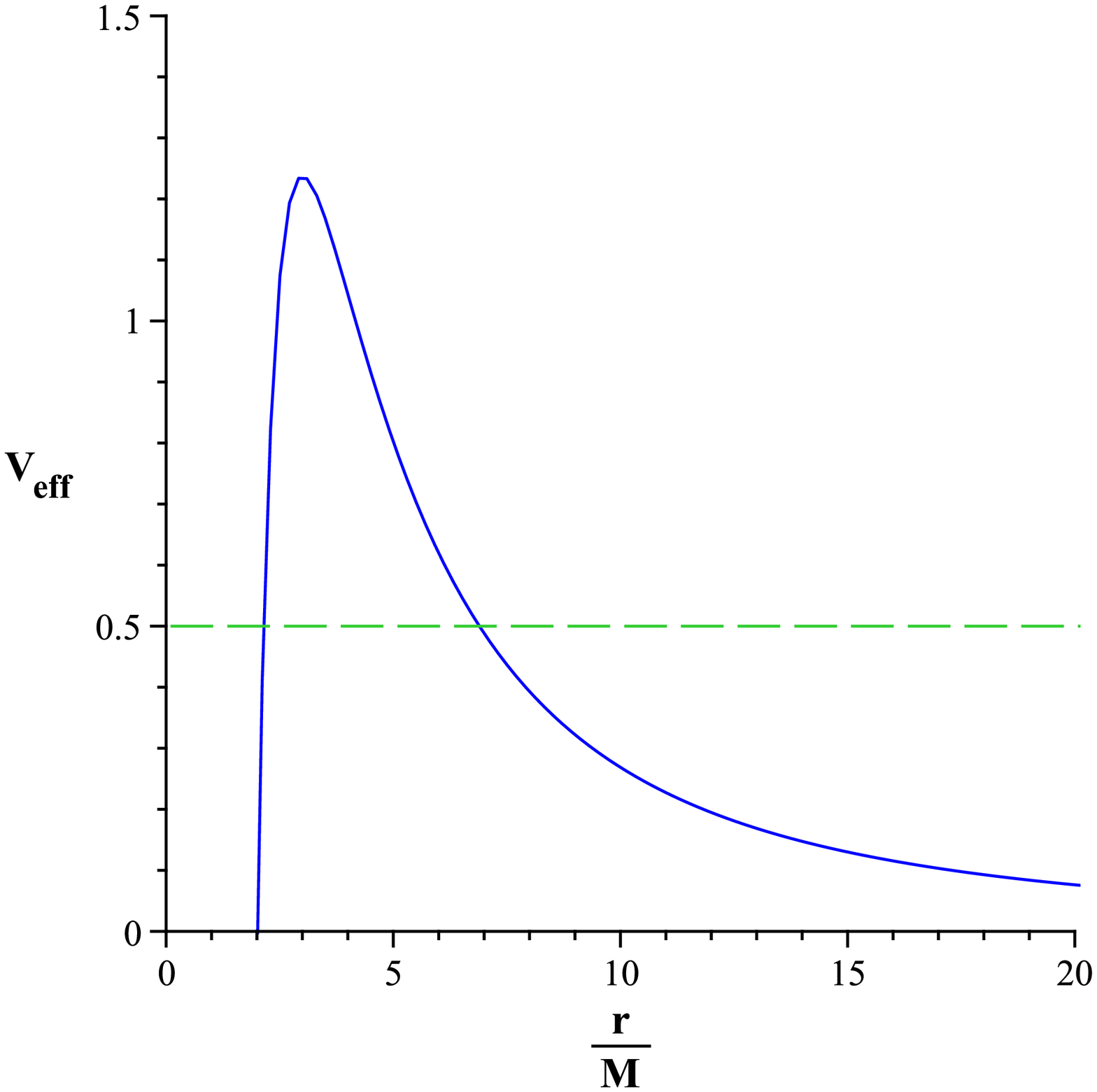}}
\caption{\label{Vnull2z}\small   
 Effective potentials for region II of parametric diagram in Fig.~(\ref{elnull}) with the parameters $\mathcal{L}=0.03, \varepsilon=0,\tilde{\beta}=1,\tilde{\gamma}=10^{-3},\tilde{k}=(\frac{1}{3})10^{-5}$. The horizontal green dashed
line represents the squared energy parameter $E^2=0.5$.
}
\end{figure}

\begin{figure}[ht]
%\centerline{\includegraphics[width=9cm]{o2gama3F.eps}}
\centerline{\includegraphics[width=9cm]{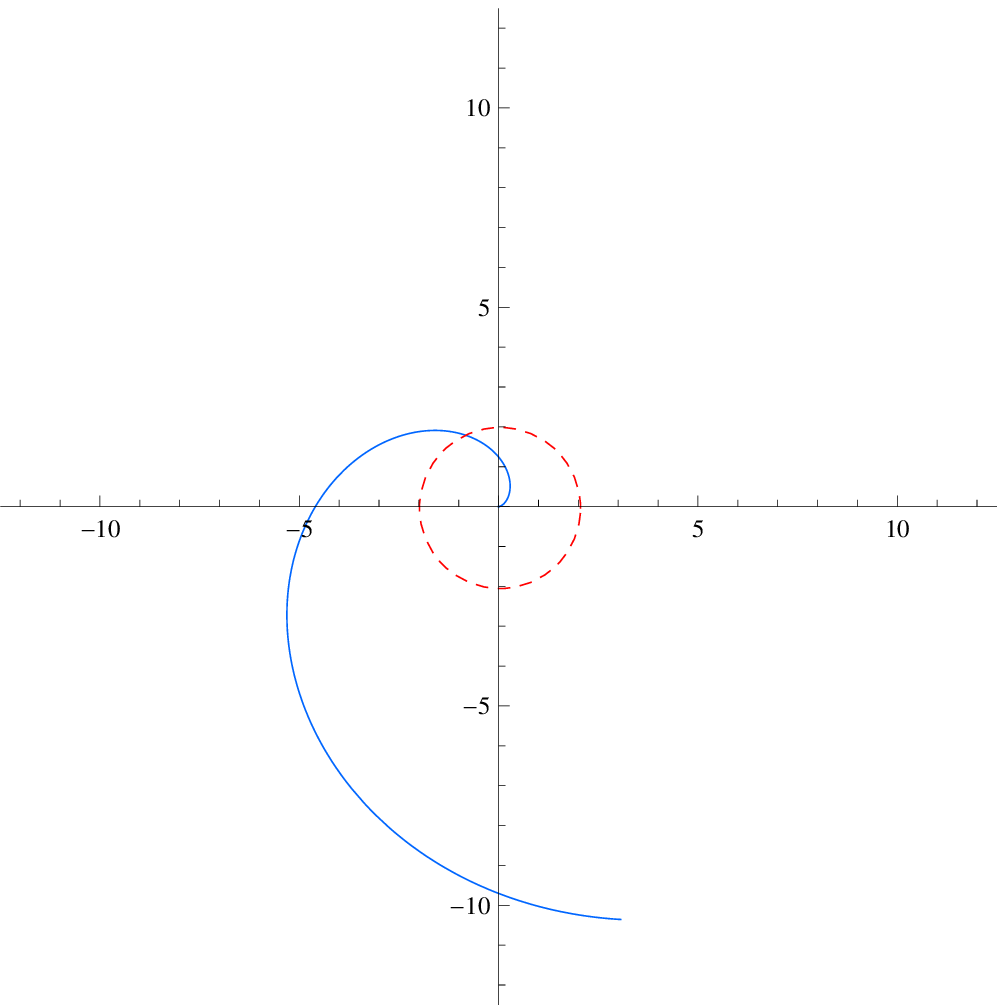}}
\caption{\label{oz0gama3}\small   
Timelike geodesics of orbit type TEO in region V of Fig.~(\ref{elnull}) with the parameters $E^2=1, \mathcal{L}=0.05, \varepsilon=1,\tilde{\beta}=1,\tilde{\gamma}=10^{-3},\tilde{k}=(\frac{1}{3})10^{-5}$.
}
\end{figure}

\begin{figure}[ht]
\centerline{\includegraphics[width=9cm]{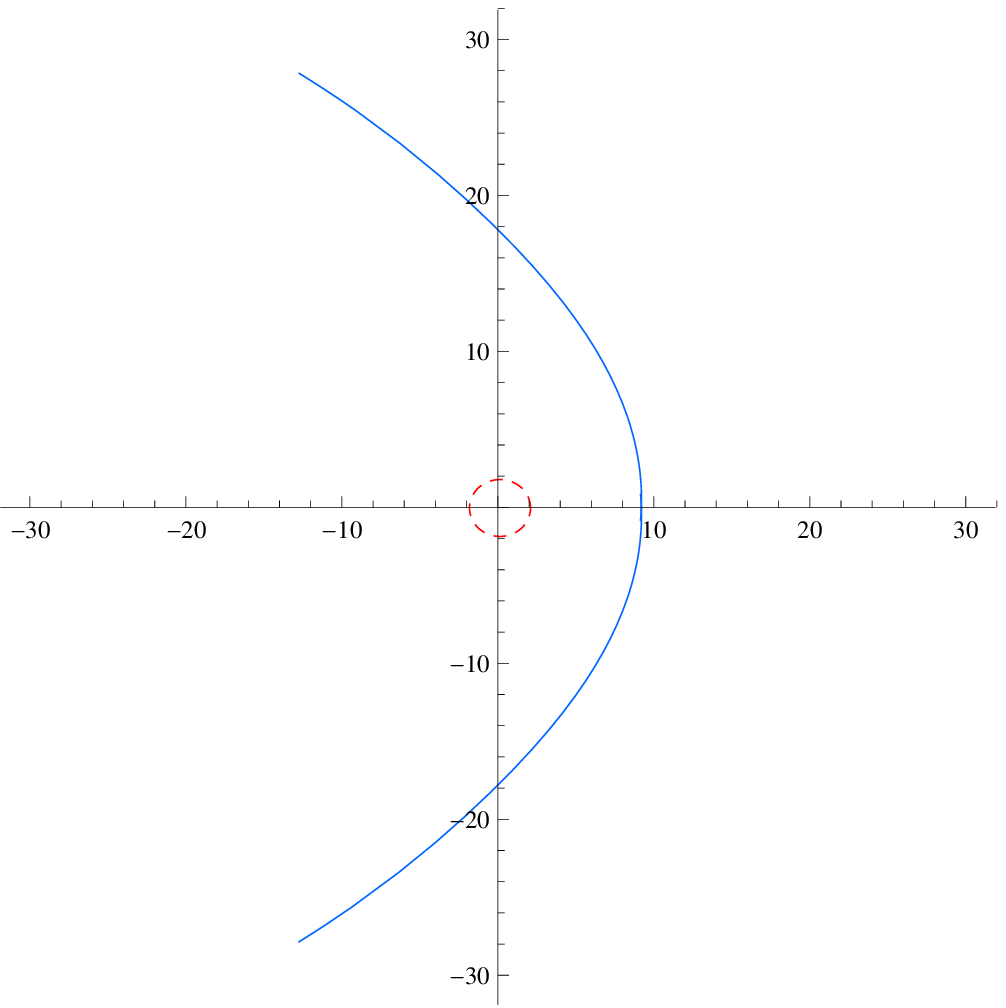}}
%\centerline{\includegraphics[width=9cm]{o2gama3E.eps}}
\caption{\label{o2gama3F}\small   
Timelike geodesics of orbit type EO in region II of Fig.~(\ref{el1}) with the parameters $E^2=1, \mathcal{L}=0.05, \varepsilon=1,\tilde{\beta}=1,\tilde{\gamma}=10^{-3},\tilde{k}=(\frac{1}{3})10^{-5}$.
}
\end{figure}

\begin{figure}[ht]
%\centerline{\includegraphics[width=9cm]{o2gama3F.eps}}
\centerline{\includegraphics[width=9cm]{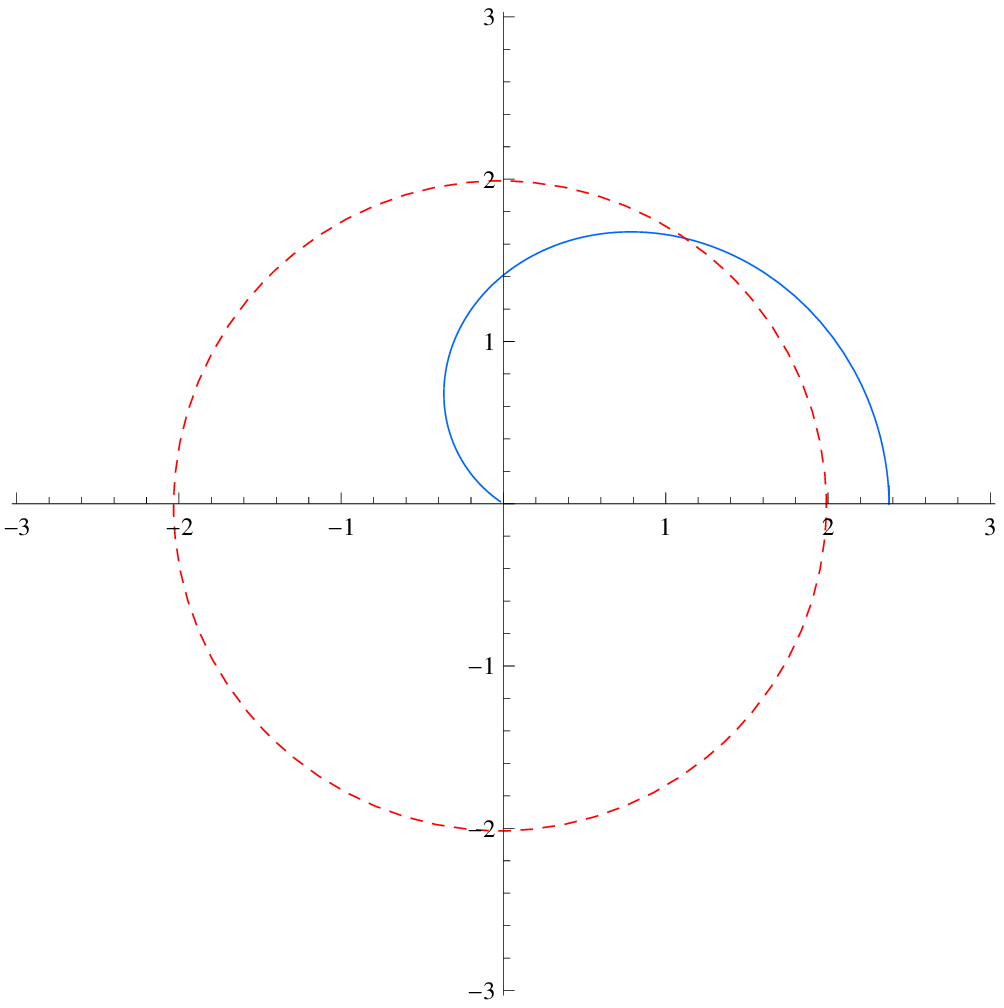}}
\caption{\label{o2gama3E}\small   
Timelike geodesics of orbit type TBO in region II of Fig.~(\ref{el1}) with the parameters $E^2=1, \mathcal{L}=0.05, \varepsilon=1,\tilde{\beta}=1,\tilde{\gamma}=10^{-3},\tilde{k}=(\frac{1}{3})10^{-5}$.
}
\end{figure}

\begin{figure}[ht]
\centerline{\includegraphics[width=8cm]{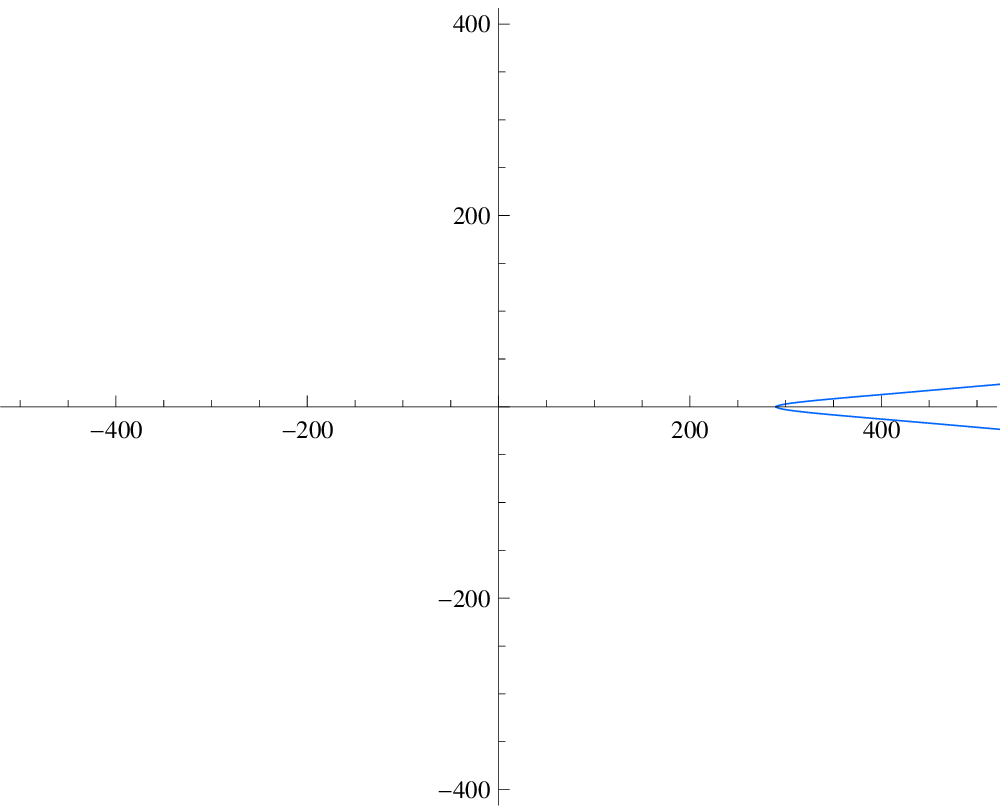}}
\caption{\label{o4zgama3fb}\small   
Timelike geodesics of orbit type EO in region IV of Fig.~(\ref{el1}) with the parameters $E^2=2, \mathcal{L}=0.001, \varepsilon=1,\tilde{\beta}=1,\tilde{\gamma}=10^{-3},J=0.1, \tilde{k}=(\frac{1}{3})10^{-5}$.
}
\end{figure}

\begin{figure}[ht]
\centerline{\includegraphics[width=8cm]{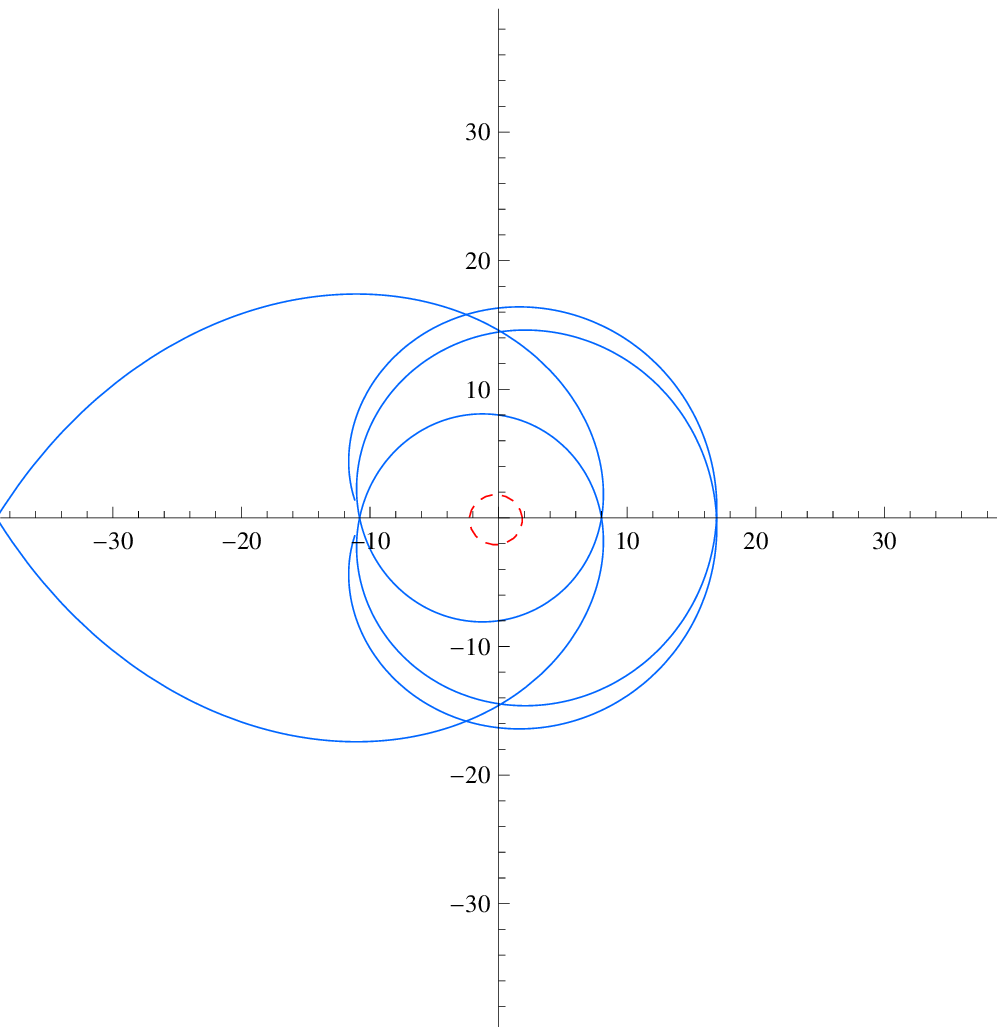}}
\caption{\label{o4zgama3bound}\small   
Timelike geodesics of orbit type BO in region IV of Fig.~(\ref{el1}) with the parameters $E^2=2, \mathcal{L}=0.05, \varepsilon=1,\tilde{\beta}=1,\tilde{\gamma}=10^{-3}, \tilde{k}=(\frac{1}{3})10^{-5}$.
}
\end{figure}

\begin{figure}[ht]
\centerline{\includegraphics[width=9cm]{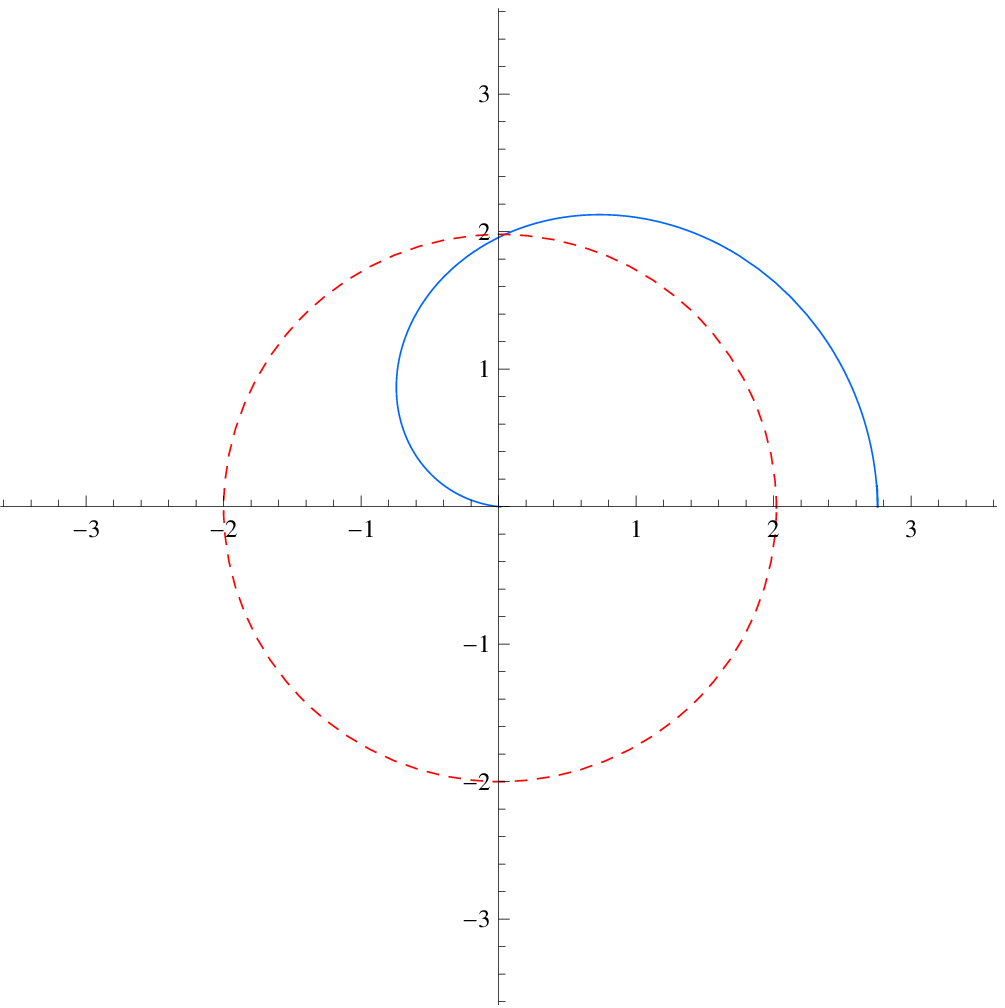}}
\caption{\label{o4zgama3EO}\small   
Timelike geodesics of orbit type TBO in region IV of Fig.~(\ref{el1}) with the parameters $E^2=1, \mathcal{L}=0.05, \varepsilon=1,\tilde{\beta}=1,\tilde{\gamma}=10^{-3}, \tilde{k}=(\frac{1}{3})10^{-5}$.
}
\end{figure}

\begin{figure}[ht]
%\centerline{\includegraphics[width=9cm]{o2gama3F.eps}}
\centerline{\includegraphics[width=9cm]{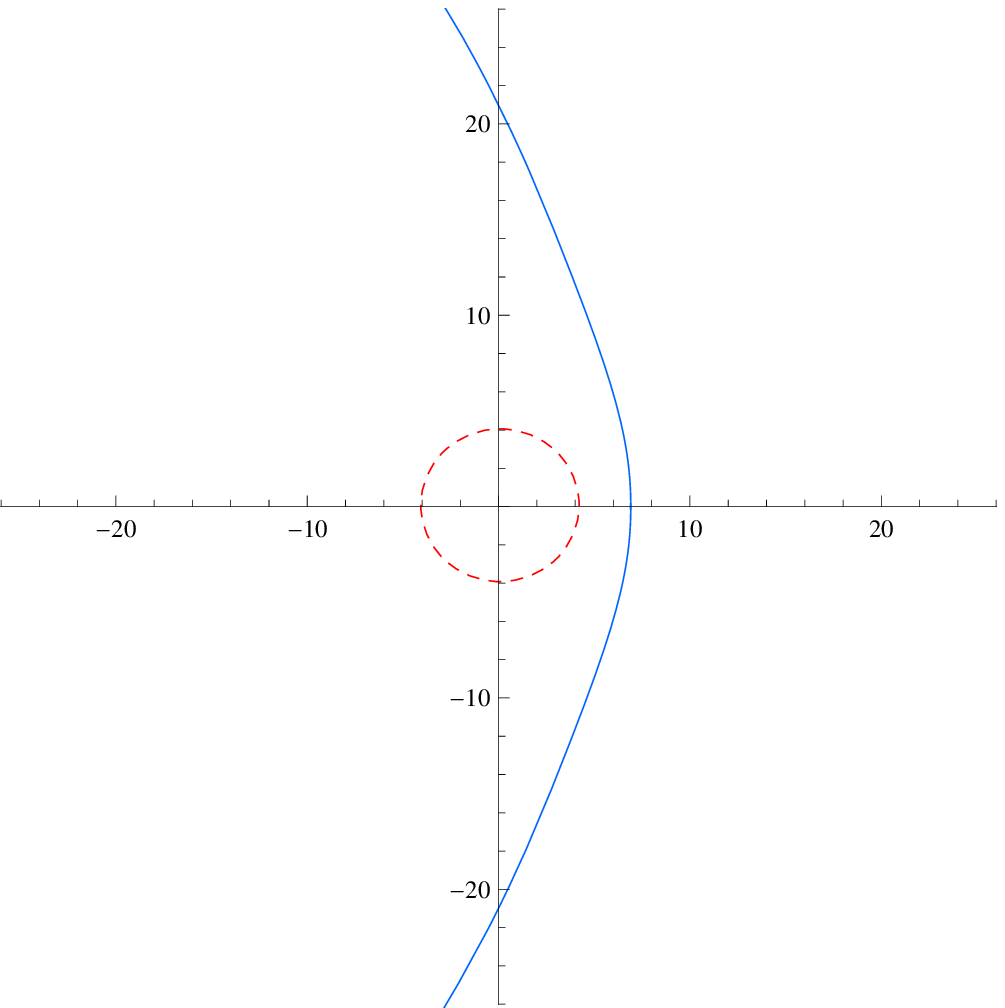}}
\caption{\label{null2zFO}\small   
Null geodesics of orbit type EO in region II of Fig.~(\ref{elnull}) with the parameters $E^2=1, \mathcal{L}=0.05, \varepsilon=0,\tilde{\beta}=1,\tilde{\gamma}=10^{-3},\tilde{k}=(\frac{1}{3})10^{-5}$.
}
\end{figure}

\begin{figure}[ht]
%\centerline{\includegraphics[width=9cm]{o2gama3F.eps}}
\centerline{\includegraphics[width=9cm]{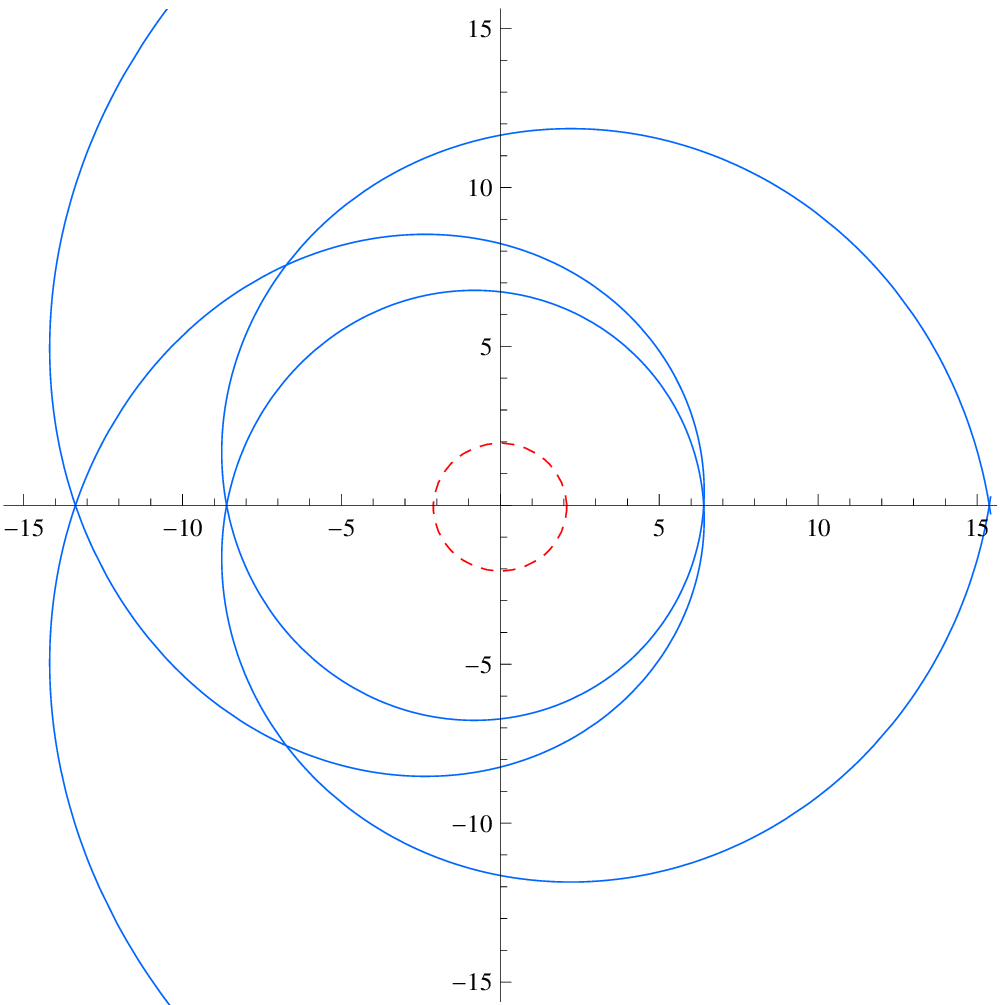}}
\caption{\label{compare1bf}\small   
Timelike geodesics of orbit type BO in region IV of Fig.~(\ref{elnull}) with the parameters $E^2=1.05, \mathcal{L}=0.055, \varepsilon=1,\tilde{\beta}=1,\tilde{\gamma}=10^{-2},\tilde{k}=(\frac{1}{3})10^{-5}$.
}   
\end{figure}

\begin{figure}[ht]
%\centerline{\includegraphics[width=9cm]{o2gama3F.eps}}
\centerline{\includegraphics[width=9cm]{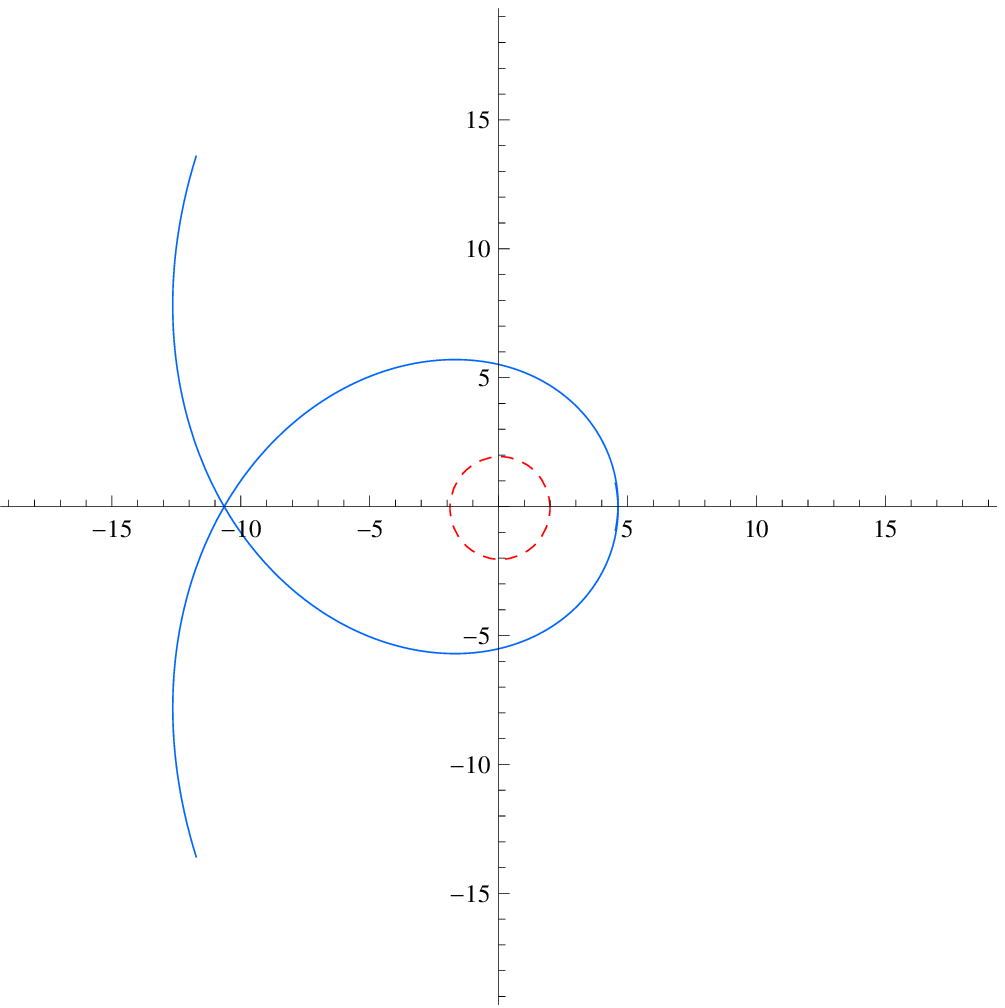}}
\caption{\label{compare1fb}\small   
Timelike geodesics of orbit type EO in region II of Fig.~(\ref{elnull}) with the parameters $E^2=1.05, \mathcal{L}=0.055, \varepsilon=1,\tilde{\beta}=1,\tilde{\gamma}=0,\tilde{k}=(\frac{1}{3})10^{-5}$.
}
\end{figure}

\begin{figure}[ht]
%\centerline{\includegraphics[width=9cm]{o2gama3F.eps}}
\centerline{\includegraphics[width=9cm]{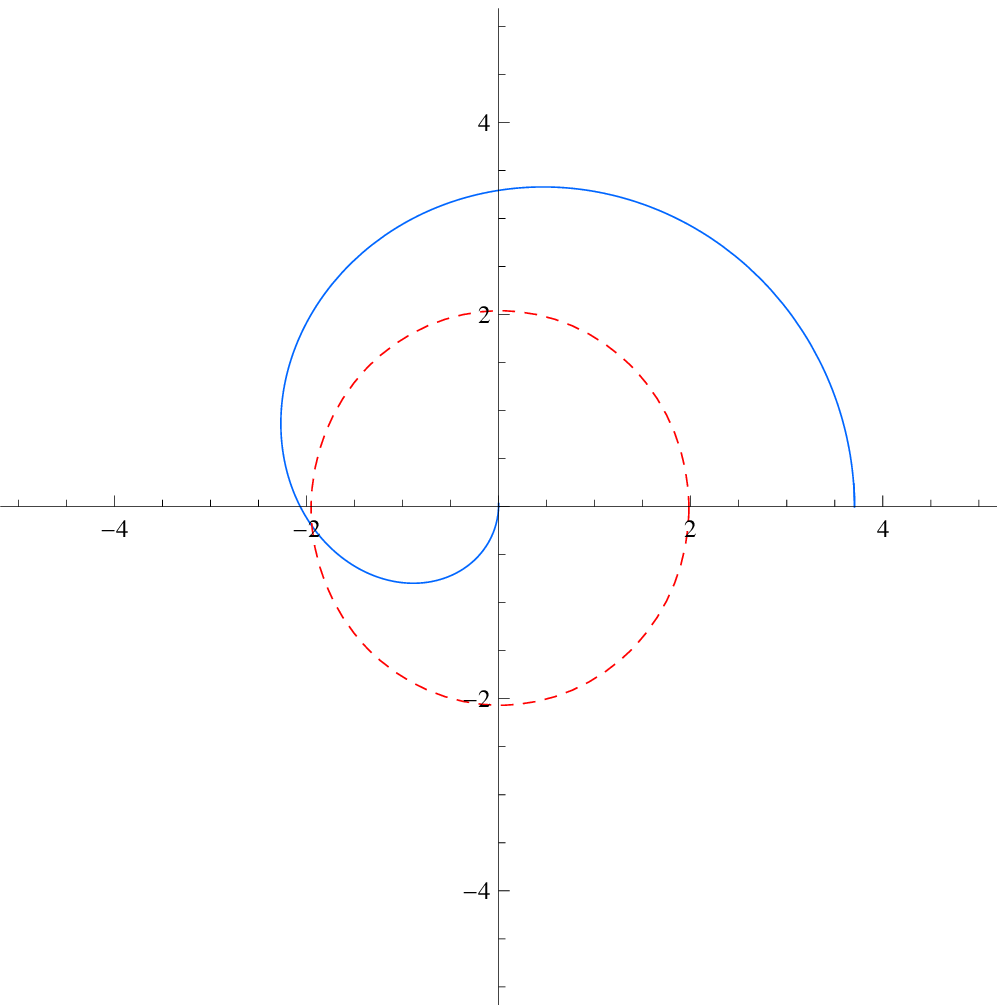}}
\caption{\label{compare2TBOF}\small   
Timelike geodesics of orbit type TBO in region II of Fig.~(\ref{elnull}) with the parameters $E^2=0.96, \mathcal{L}=0.07, \varepsilon=1,\tilde{\beta}=1,\tilde{\gamma}=10^{-2},\tilde{k}=(\frac{1}{3})10^{-5}$.
}
\end{figure}

\begin{figure}[ht]
%\centerline{\includegraphics[width=9cm]{o2gama3F.eps}}
\centerline{\includegraphics[width=9cm]{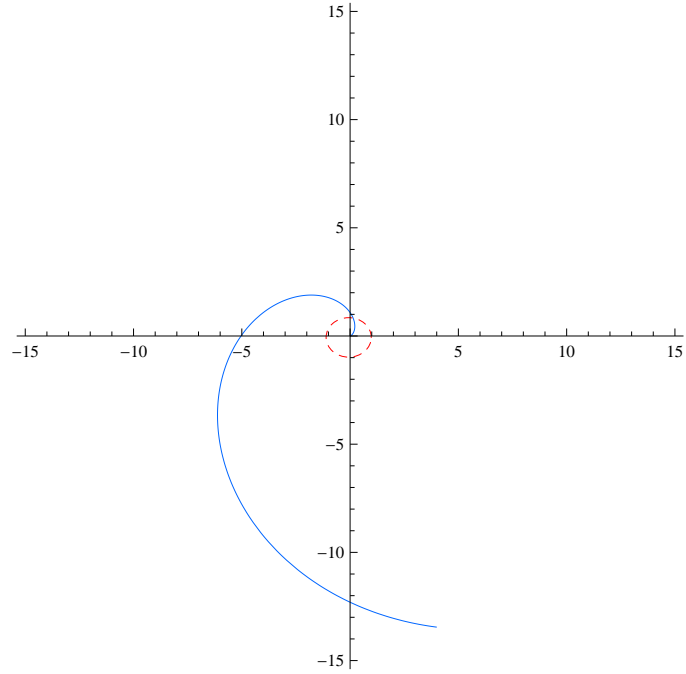}}
\caption{\label{compare2FTBO}\small   
Timelike geodesics of orbit type TEO in region II of Fig.~(\ref{elnull}) with the parameters $E^2=0.96, \mathcal{L}=0.07, \varepsilon=1,\tilde{\beta}=1,\tilde{\gamma}=0,\tilde{k}=(\frac{1}{3})10^{-5}$.
}
\end{figure}

% The real and positive zeros of $R^*(\tilde{r})$ are extremal values of the geodesic motion and its zeros are the turning points of the geodesics.
By solving the system of equations $R(\tilde{r})=0$ and $\frac{dR(\tilde{r})}{d\tilde{r}}=0$ for $\epsilon=1$ we can solve the equations for $E^2$ and $\mathcal{L}$
\begin{equation}\label{EL}
\mathcal{L}=-\frac{(\tilde{r}-3\tilde{\beta})(\tilde{\gamma} \tilde{r}-3\tilde{\beta}\tilde{\gamma} +2)}{\tilde{r}^2(2\tilde{r}^3\tilde{k}-\tilde{\gamma} \tilde{r}^2-2\tilde{\beta} +3\tilde{\beta}^2\tilde{\gamma})},\quad E^2=\frac{2(-\tilde{r}-\tilde{r}^2\tilde{\gamma} +2\tilde{\beta} +\tilde{k}\tilde{r}^3+3\tilde{\beta}\tilde{\gamma} \tilde{r}-3\tilde{\beta}^2\tilde{\gamma})^2}{\tilde{r}(\tilde{r}-3\tilde{\beta})(\tilde{\gamma} \tilde{r}-3\tilde{\beta}\tilde{\gamma} +2)},
\end{equation}
 
and for $\epsilon=0$

\begin{equation}\label{EL0}
\mathcal{L}=-\frac{1}{27}\frac{27\tilde{k}\tilde{\beta}^2 -3\tilde{\beta}\tilde{\gamma} -1}{\tilde{\beta}^2 E^2}.
\end{equation}
Using Eqs.~(\ref{EL}) and (\ref{EL0}), we plot the parametric diagram 
 for test particle $(\epsilon=1)$ and light $(\epsilon=0)$, in conformal spacetime, which is shown in Figs.~\ref{el1}--\ref{elgama2}. 
 %The parameters  In Figs.(\ref{el1},\ref{elnull},\ref{elgama2}), are $\tilde{k}=\frac{10^{-5}}{3},\tilde{\gamma}=0, \tilde{\beta}=1$. We study the results for $\tilde{k}=\frac{10^{-5}}{3},\tilde{\gamma}=0, \tilde{\beta}=1$ that in this case Eq.~(\ref{E},\ref{L}), for $\epsilon=1$, 
 For $\gamma=0$, Eqs.~(\ref{EL}) and (\ref{EL0}), reduce to Schwarzschild-de Sitter expressions \cite{Hackmann:2008zza} 
 
 \begin{equation}\label{ELgama01}
\epsilon=1 \longrightarrow \left\{
\begin{array}{rl}
E^2=\frac{(\tilde{k}\tilde{r}^3-\tilde{r}+2)^2}{\tilde{r}(\tilde{r}-3)},  \\
\mathcal{L}=\frac{-(\tilde{r}-3)}{\tilde{r}^2(\Lambda \tilde{r}^3-1)},   \\
\end{array} \right.
\end{equation}
 
 and
 
 \begin{equation}\label{ELgama00}
 \epsilon=0 \longrightarrow  \mathcal{L}=\frac{1}{E^2}(\frac{1}{27}-\tilde{k}).
 \end{equation}
 
and for $\tilde{k}=0, \tilde{\gamma}=0$, reduce to Schwarzschild expressions

 \begin{equation}\label{ELk1}
\epsilon=1 \longrightarrow \left\{
\begin{array}{rl}
 E^2=\frac{(\tilde{r}-2)^2 }{\tilde{r}(\tilde{r}-3)},  \\
\mathcal{L}=\frac{(\tilde{r}-3)}{\tilde{r}^2},   \\
\end{array} \right.
\end{equation}

 \begin{equation}\label{ELk0}
\epsilon=0 \longrightarrow \mathcal{L}=\frac{1}{27E^2}.
 \end{equation}
The parametric $\mathcal{L}$-$E^2$ diagram for Schwarzschild-de Sitter and Schwarzschild spacetime, are shown in Figs.~\ref{elshdesiter} and \ref{elsh}.
%which is ploted by Eqs.(\ref{ELgama01}) and (\ref{ELk1}).
 Also, overview of possible regions for these spacetimes (CG, Schwarzschild spacetime and Schwarzschild-de Sitter spacetime), are shown in Table.~\ref{tab:cyl.orbitsh}. 
The zeros number of polynomials $R(r)$, varies with different parameters $\tilde{k}, \tilde{\beta}, \epsilon, E^2$ and $\mathcal{L}$, so, the regions for various parameters are different. In Fig.~(\ref{el1}), three region can be identified.
It can be seen from $\mathcal{L}$-$E^2$ diagrams (Figs.~\ref{el1}--\ref{elgama2}) and Table.~\ref{tab:cyl.orbitsh} , for CG, region II, IV and V are appeared, and for   
 null geodesics, there is only two regions (IV and II). We plot the effective potential for different regions in conformal gravity (see Figs.~\ref{V1o0}--\ref{Vnull2z}).  
% the result of Eq.~(\ref{E},\ref{L}) are shown for test particle $\epsilon=1$ and for positive cosmolgy constant $(\Lambda>0)$. Three regions can be identified in this diagram
In region V, the polynomials $R(r)$, does not have any zeroes and $R(\tilde{r})>0$ for positive $\tilde{r}$, Possible orbit type for this region is terminating escape orbits (TEO), (see Fig.~\ref{oz0gama3}). In region II, there are two positive real zeros for $R(r)$ and possible orbit types are escape orbit (EO) and terminating bound orbit (TBO) (see Figs.~\ref{o2gama3F} and \ref{o2gama3E}), also in region III, $R^*(r)$ has 4 positive real zeros which its possible orbits are escape orbit (EO), bound (BO), and terminating bound orbit (TBO), (see Figs.~\ref{o4zgama3fb}--\ref{o4zgama3EO}). Moreover the deflection of light is shown in Fig.~\ref{null2zFO}. By comparing the geodesic motion by $\gamma\neq 0$ of test particles with the case $\gamma=0$ in  Figs.~\ref{compare1bf}--\ref{compare2FTBO}, it is clear that for $\gamma\neq 0$, regions are a little
shifted to $\gamma=0$, and also some of the types of orbits are influenced, which can be seen in Figs.~\ref{compare1bf}--\ref{compare2FTBO}.

     \begin{table}[h]
\begin{center}
\begin{tabular}{|l|l|c|l|}
%{|lccll|}
\hline
region & pos.zeros  & range of $\tilde{r}$ & orbit \\
\hline\hline

I & 1 & 
\begin{pspicture}(-3,-0.2)(2.2,0.2)%\psgrid
\psline[linewidth=0.5pt]{->}(-2.5,0)(1.5,0)
\psline[linewidth=0.5pt](-2.5,-0.2)(-2.5,0.2)
\psline[linewidth=0.5pt,doubleline=true](-2,-0.2)(-2,0.2)
%\psline[linewidth=0.5pt,doubleline=true](-0.5,-0.2)(-0.5,0.2)
%\psline[linewidth=0.5pt,doubleline=true](1,-0.2)(1,0.2)
\psline[linewidth=1.2pt]{-*}(-2.5,0)(-1.5,0)
%\psline[linewidth=1.2pt]{*-}(-0.1,0)(1.5,0)
\end{pspicture}
  & TBO, EO
\\ \hline
II & 2 & 
\begin{pspicture}(-3,-0.2)(2.2,0.2)%\psgrid
\psline[linewidth=0.5pt]{->}(-2.5,0)(1.5,0)
\psline[linewidth=0.5pt](-2.5,-0.2)(-2.5,0.2)
\psline[linewidth=0.5pt,doubleline=true](-2,-0.2)(-2,0.2)
%\psline[linewidth=0.5pt,doubleline=true](-0.5,-0.2)(-0.5,0.2)
%\psline[linewidth=0.5pt,doubleline=true](1,-0.2)(1,0.2)
\psline[linewidth=1.2pt]{-*}(-2.5,0)(-1.5,0)
\psline[linewidth=1.2pt]{*-}(-0.1,0)(1.5,0)
\end{pspicture}
  & TBO, EO
\\ \hline
III  & 3 &
\begin{pspicture}(-3,-0.2)(2.2,0.2)%\psgrid
\psline[linewidth=0.5pt]{->}(-2.5,0)(1.5,0)
\psline[linewidth=0.5pt](-2.5,-0.2)(-2.5,0.2)
\psline[linewidth=0.5pt,doubleline=true](-2,-0.2)(-2,0.2)
%\psline[linewidth=0.5pt,doubleline=true](-0.5,-0.2)(-0.5,0.2)
%\psline[linewidth=0.5pt,doubleline=true](1,-0.2)(1,0.2)
%\psline[linewidth=1.2pt]{-*}(-4,0)(-3,0)
\psline[linewidth=1.2pt]{*-*}(-0.5,0)(0.5,0)
\psline[linewidth=1.2pt]{-*}(-2.5,0)(-1.5,0)
%\psline[linewidth=1.2pt]{*-}(0.7,0)(1.5,0)
\end{pspicture}
& EO,TO
\\ \hline
IV  & 4 &
\begin{pspicture}(-3,-0.2)(2.2,0.2)%\psgrid
\psline[linewidth=0.5pt]{->}(-2.5,0)(1.5,0)
\psline[linewidth=0.5pt](-2.5,-0.2)(-2.5,0.2)
\psline[linewidth=0.5pt,doubleline=true](-2,-0.2)(-2,0.2)
%\psline[linewidth=0.5pt,doubleline=true](-0.5,-0.2)(-0.5,0.2)
%\psline[linewidth=0.5pt,doubleline=true](1,-0.2)(1,0.2)
%\psline[linewidth=1.2pt]{-*}(-4,0)(-3,0)
\psline[linewidth=1.2pt]{*-*}(-0.75,0)(0,0)
\psline[linewidth=1.2pt]{-*}(-2.5,0)(-1.5,0)
\psline[linewidth=1.2pt]{*-}(0.7,0)(1.5,0)
\end{pspicture}
& TBO, BO, FO 
\\ \hline
%\\ \hline
V & 0 & 
\begin{pspicture}(-3,-0.2)(2.2,0.2)%\psgrid
%\psline[linewidth=0.5pt]{->}(-2.5,0)(1.5,0)
%\psline[linewidth=0.5pt](-2.5,-0.2)(-2.5,0.2)
%\psline[linewidth=0.5pt,doubleline=true](-0.5,-0.2)(-0.5,0.2)
%\psline[linewidth=0.5pt,doubleline=true](1,-0.2)(1,0.2)
%\psline[linewidth=1.2pt]{-*}(-4,0)(-3,0)
%\psline[linewidth=1.2pt]{*-*}(-1,0)(1.5,0)
%\psline[linewidth=1.2pt]{-}(-2.5,0)(1.5,0)
\psline[linewidth=0.5pt]{->}(-2.5,0)(1.5,0)
\psline[linewidth=0.5pt](-2.5,-0.2)(-2.5,0.2)
\psline[linewidth=0.5pt,doubleline=true](-2,-0.2)(-2,0.2)
%\psline[linewidth=0.5pt,doubleline=true](-0.5,-0.2)(-0.5,0.2)
%\psline[linewidth=0.5pt,doubleline=true](1,-0.2)(1,0.2)
%\psline[linewidth=1.2pt]{-*}(-4,0)(-3,0)
%\psline[linewidth=1.2pt]{*-*}(-0.5,0)(0.5,0)
\psline[linewidth=1.2pt]{-}(-2.5,0)(1.5,0)
%\psline[linewidth=1.2pt]{*-}(0.7,0)(1.5,0)
\end{pspicture}
  & TEO

\\ \hline\hline
\end{tabular}
\caption{Types of orbits of Schwarzschild spacetime in CG. The range of the orbits is represented by thick lines. The dots show the turning points of the orbits.
% The positions of the horizons are marked by a vertical double line.
 The single vertical line indicates $\tilde{r}=0$. The event horizon, is marked by a double vertical line.}
\label{tab:cyl.orbitsh}
\end{center}
\end{table}

% \begin{figure}[ht]
%\centerline{\includegraphics[width=8cm]{ELanti.eps}}
%\caption{\label{elanti}\small   
% Parametric $\mathcal{L}$-$E^2$-diagram with the parameters $\varepsilon=1,\tilde{\beta}=1,\tilde{\gamma}=10^{-3}, \tilde{k}=-\frac{10^{-5}}{3}$. There are three different regions. $R^{*}$ has one zero in region I, two zeros in region II, none zero in region III and four zeores in region IV for conformal gravity by negative cosmology constant.
%}
%\end{figure}


\clearpage

\begin{figure}[ht]
%\centerline{\includegraphics[width=9cm]{o2gama3F.eps}}
\centerline{\includegraphics[width=9cm]{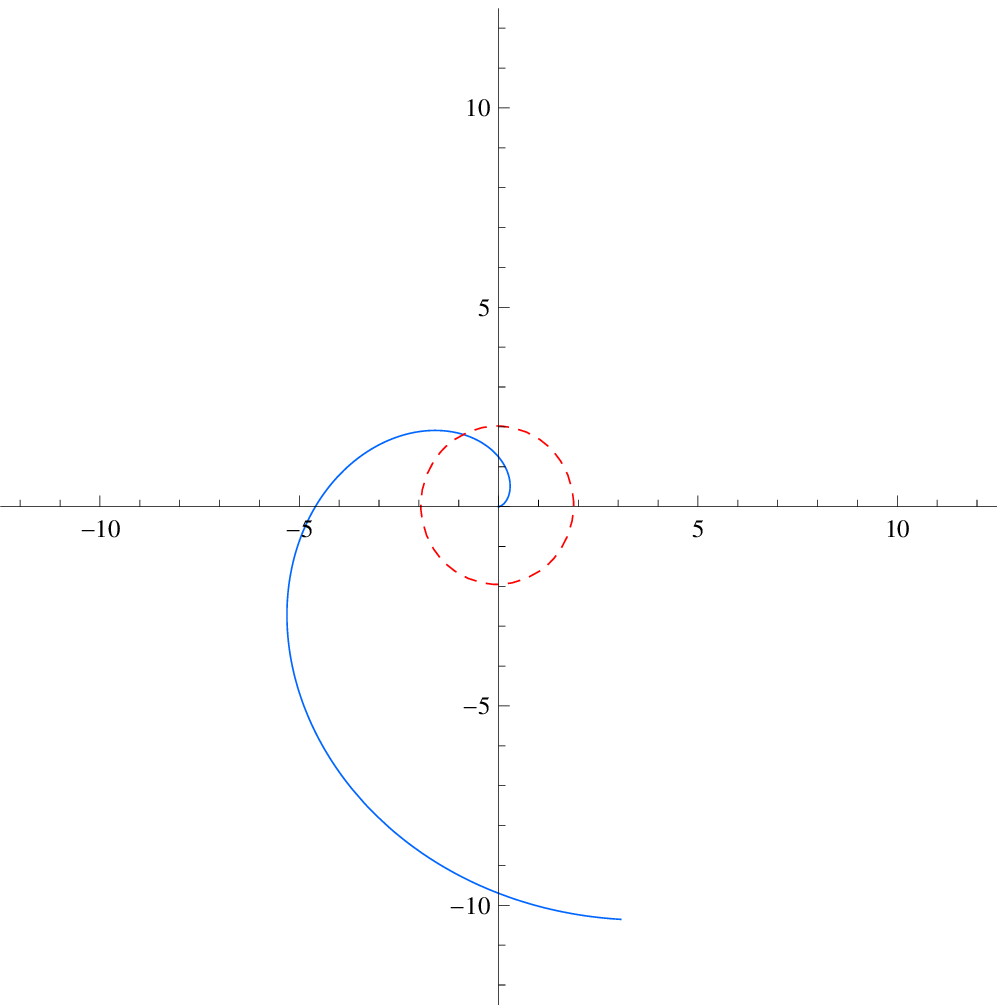}}
\caption{\label{null0z}\small   
Null geodesics of orbit type TEO in region V of Fig.~(\ref{elnull}) with the parameters $E^2=1, \mathcal{L}=0.05, \varepsilon=0,\tilde{\beta}=1,\tilde{\gamma}=10^{-3},\tilde{k}=(\frac{1}{3})10^{-5}$.
}
\end{figure}

\begin{figure}[ht]
%\centerline{\includegraphics[width=9cm]{o2gama3F.eps}}
\centerline{\includegraphics[width=9cm]{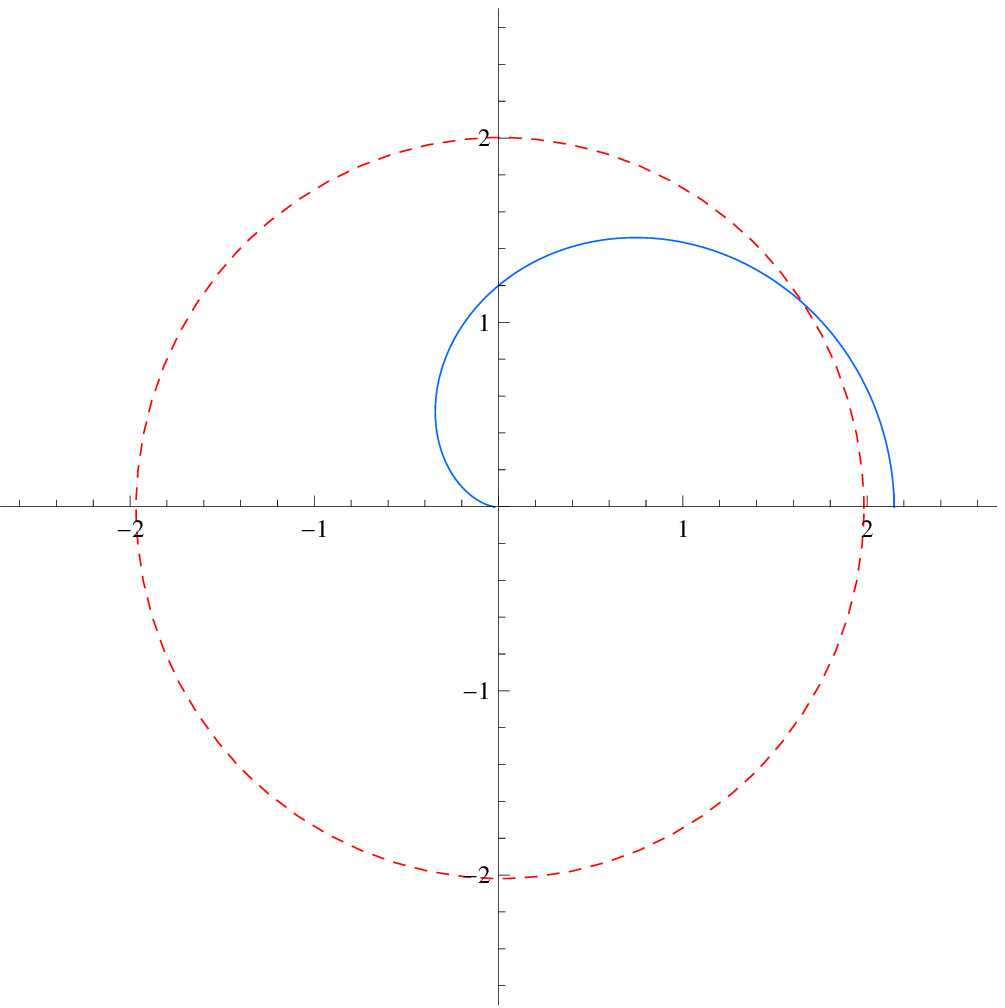}}
\caption{\label{null2zEO}\small   
Null geodesics of orbit type TBO in region II of Fig.~(\ref{elnull}) with the parameters $E^2=1, \mathcal{L}=0.05, \varepsilon=0,\tilde{\beta}=1,\tilde{\gamma}=10^{-3},\tilde{k}=(\frac{1}{3})10^{-5}$.
}
\end{figure}

\clearpage

\section{CONCLUSIONS}\label{C}
In this paper, we studied the analytical solution and possible orbits for test particles and light rays in spherical conformal spacetime. We used Weierstrass elliptic functions, and derivatives of Kleinian sigma functions. Studying the possible orbits in each region of parametric diagram, we found out that there are three regions for the timelike geodesics and two regions for the geodesics of light. Also we compare the results of conformal spacetime with Schwarzschild de Sitter which was studyed in Ref.\cite{Hackmann:2008zz}.  
Using analytic solutions we calculated the light deflection of escape orbit (EO). 
The results of this paper are a useful tool to investigate the periastron shift of bound orbits, the deflection angle and the Lense-Thirring effect.
 For future it would be interesting to study equations of motion in the charged and rotating version of this black hole spacetime.

\bibliographystyle{amsplain}

\end{document}